# Thermally Induced Structural Evolution and Nanoscale Interfacial Dynamics in Layered Metal Chalcogenides

*Parivash Moradifar[1,2*], Tao Wang[3,4], Nadire Nayir[3,4,5], Tiva Sharifi[6,7], Ke Wang[1], Pulickel Ajayan[6], Adri C.T. van Duin[3,4], Nasim Alem[1*]*

[1]Department of Materials Science and Engineering, Materials Research Institute, The Pennsylvania State University, University Park, PA 16802, USA

[2]Department of Materials Science and Engineering, Stanford University, Stanford, California 94305, USA

[3]Department of Mechanical Engineering, The Pennsylvania State University, University Park, PA, 16802, United States

[4]2D Crystal Consortium (2DCC), Materials Research Institute (MRI), The Pennsylvania State University, University Park, PA 16802, USA

[5]Department of Physics, Karamanoglu Mehmetbey University, Karaman, 70000, Turkey[c]

[6]Department of Materials Science and Nano Engineering, Rice University, Houston, TX 77005, USA

[7]Department of Physics, Umeå University, Umeå 90187, Sweden

*Correspondence and requests for material should be addressed to Parivash Moradifar (email: pmoradi@stanford.edu) and Nasim Alem (email: nua10@psu.edu).

## Abstract

Layered chalcogenides including $Bi_2Te_3$, $Sb_2Te_3$, Bi-Sb-Te ternary alloys and heterostructures are known as great thermoelectric, topological insulators and recently highlighted as plasmonic building blocks beyond noble metals. Here, we conduct a joint *in situ* transmission electron microscopy (*in situ* TEM) and density functional theory (DFT) calculations to investigate the temperature dependent nanoscale dynamics, interfacial properties and further identifying the role of native defects and edge configurations in anisotropic sublimations of $Bi_2Te_3$-$Sb_2Te_3$ in-plane heterostructure and $Sb_{2-x}Bi_xTe_3$ alloy. Structural dynamics including edge evolution, formation, expansion, and coalescence of thermally induced polygonal nanopores are reported. The nanopores appear to be initiated by preferential dissociation of chalcogenide species (Te) from the center, heterointerface and edges in the heterostructure and only from the outer edges in the alloy counterpart. This results in a reduced thermal stability and significantly different sublimation pathways of the heterostructure. Furthermore, triangular and quasi hexagonal configurations are observed to be the dominant nanopores configurations in the heterostructure. Additionally, our DFT calculations provide a mechanistic understanding on the role of native defects and edge formation energies, revealing the antisite defects $Te_{Bi}$ to be the dominant native defect in a Te-rich condition and playing a key role on the defects assisted sublimation. These findings significantly impact our understanding of controlling the nanoscale sublimation dynamics and can ultimately assist us in designing tunable low-dimensional chalcogenides.

## Introduction

V-VI group chalcogenides such as bismuth telluride ($Bi_2Te_3$) and antimony telluride ($Sb_2Te_3$) are known for their outstanding thermoelectric (TE) and topological insulator (TI) properties (1). Exhibiting gapless metallic surface states, they have recently garnered significant attention in the field of nanophotonics (2) (3), and have also been proposed as new emerging plasmonic building blocks beyond noble metals (Ag, Au) (4) (5). In particular, providing multiple edges, large surface areas and strong light-matter interaction in comparison with their bulk counterparts, they are particularly advantageous for plasmonic and energy applications (6), optical switching and catalysts for energy conversion applications (3) (7).

To unlock the full potential of low-dimensional chalcogenides, it is imperative to design tunable platforms at the atomic level and to be able to manipulate the structure with atomic precision. This is because local microstructural heterogeneities, i.e. vacancies, antisites, dopants, interfaces, and edges, can significantly alter the macroscopic properties. Therefore, understanding and controlling the atomic structure is the key to ultimately design tunable low-dimensional crystals. Microstructural heterogeneities can be considered as intrinsic (pre-existing) or extrinsic (externally induced) defects. While intrinsic defects naturally exist in the crystal, extrinsic defects can be created by external stimuli such as irradiation (electrons, ions or laser pulses) or stress (thermal, mechanical etc.), and further modulate the electronic, optical, and optoelectronic properties of the crystal. As an example, thermally induced polygonal defects are proposed as a tool for tuning and selective enhancement of the plasmonic response in two-dimensional (2D) $Bi_2Te_3$-$Sb_2Te_3$ in-plane heterostructures over a broad spectral range (8). In addition, ion-irradiation induced vacancy clusters is correlated with the formation of characteristic charged and neutral excitons in low-dimensional $MoS_2$ and $WSe_2$ (9) (10) (11) as well as binding excitons and trions (9). Similarly, extrinsic defects can modify the electronic properties such as the band gap in transition metal dichalcogenides (TMDCs) (12) (13) (14).

Scanning/transmission electron microscopy (S/TEM) has enabled real-time *in situ* observation of the crystal structure, defect dynamics, and transformation pathways of various classes of materials under external stimuli in various environments with near atomic precision. It can also be used to create or manipulate defects in real-time (15) (16) (17). There have been a few reports on exploring the phase stabilities of Ge-Sb-Te alloy (18), and sublimation of $Bi_2Te_3$ and $Bi_2Se_3$ nanocrystals under thermal stress via *in situ* TEM (19) (20) (21). In a recent study, the impact of copper intercalation on structural transformation of $Bi_2Te_3$ has been investigated (22) (23). However, the role of alloying species and presence of a heterointerface on the sublimation pathways and structural dynamics of these layered metal chalcogenides are yet to be explored.

In this work, using *in situ* TEM equipped with a heating stage, we assess the high temperature structural dynamics, compositional stabilities, and real time monitoring of thermally induced defects in hexagonal $Bi_2Te_3$-$Sb_2Te_3$ in-plane heterostructures. The results are compared with the ternary alloy counterpart of $Bi_2Te_3$ and $Sb_2Te_3$, $Sb_{2-x}Bi_xTe_3$, to further identify how the presence of a physical heterointerface and alloying elements can alter the sublimation pathways during *in situ* TEM annealing. In this study, the hexagonal platelets of $Bi_2Te_3$-$Sb_2Te_3$ in-plane heterostructures and $Sb_{2-x}Bi_xTe_3$ alloys are prepared through a two-step solvothermal method based on a Te-seeded growth (Figure S1) (6) (24).

The experimental findings are further combined with density functional theory (DFT) calculations to gain atomic insight into the underlying mechanism and understand the potential thermodynamic pathways responsible for the preferential nanopore nucleation and anisotropic sublimation in $Bi_2Te_3$-$Sb_2Te_3$ heterostructures and $Sb_{2-x}Bi_xTe_3$ alloys. Based on our *in situ* S/TEM observations and DFT calculations, we identify the preferential edge configurations for sublimation and the most energetically favorable native defects that are responsible for the polygonal nanopore formation and the anisotropic sublimation. Understanding the physics behind the formation, growth and transformation of thermally induced defects is of great significance to further integrate defects as a tuning toolkit for a wide range of functional properties (13) (25). Additionally, a deep insight into the atomistic mechanisms governing the sublimation is of great importance in guiding the field of defect-assisted optical properties and band gap tuning and ultimately designing tunable low-dimensional chalcogenides.

## Results and Discussion

$Bi_2Te_3$ and $Sb_2Te_3$ both share a rhombohedral crystal structure with space group of $R\bar{3}m$ (Figure S2). Having a layered microstructure, $Bi_2Te_3$ and $Sb_2Te_3$ consist of alternative hexagonal monoatomic crystal planes of Bi and Te atoms that are arranged with ABC ordering along the c axis. Five monoatomic layers with the quint substructure of $Te_1$-Bi-$Te_2$-Bi-$Te_1$ form a tightly bound quintuple (QL) sheets, that is charge-neutral with the thickness of 1nm. The layered structure of $Bi_2Te_3$ features strong intralayer covalent bonds between Bi and Te while experiencing weak interlayer van der Waals forces (3) (24) (26) (27).

Figure 1 shows TEM structural analysis on 2D $Bi_2Te_3$-$Sb_2Te_3$ in-plane heterostructures confirming the single crystallinity of the hexagonal nanocrystals. Bright field (BF)-TEM of 2D $Bi_2Te_3$-$Sb_2Te_3$ in-plane heterostructure on a lacey carbon grid (Figure 1a) reveals a ripple like pattern indicating the existence of bent contours in this structure. The bent contours potentially originate from the local and inhomogeneous distribution of elastic strain as previously observed in solvothermal processed $Bi_2Te_3$ nanoplates (27) (28). Here, the presence of the heterointerfaces is also expected to further contribute to the formation of bent contours in the flakes. According to Figures 1a and 1b, bent contours initiate from the proximity of Te nanorod at the center of the flake.

Electron diffraction pattern (EDP) of the entire flake along the [0001] direction shows the 2D in-plane heterostructure is single crystalline (Figure 1c) with $Bi_2Te_3$ and $Sb_2Te_3$ diffraction spots overlapped. The extra spots marked with orange circles are associated with the Te nanorod from the same zone axis at the center of the flake. This observation is consistent with the simulated EDPs for Te nanorod, $Bi_2Te_3$, and $Sb_2Te_3$ structures (presented in Figures S3 and S4). However, due to the small lattice mismatch between $Bi_2Te_3$ and $Sb_2Te_3$, the spots slightly overlap and look elongated in both the simulations and the experimental observations in Figure 1c. The lattice parameters for $Bi_2Te_3$ and $Sb_2Te_3$ are $a_{Bi2Te3}= b_{Bi2Te3}= 4.384$ Å, $c_{Bi2Te3}= 30.487$ Å, $a_{Sb2Te3}= b_{Sb2Te3}= 4.264$ Å, $c_{Sb2Te3} = 30.458$ Å (29) and are listed in Table 1). The lattice mismatch between $Bi_2Te_3$ and $Sb_2Te_3$ is about 2.5%. The lattice mismatch between $Bi_2Te_3$ and $Sb_2Te_3$ (in x-y lattice plane) is expected to minimize the strain energy through the presence of compressive or tensile elastic in-plane strain (29) (30).

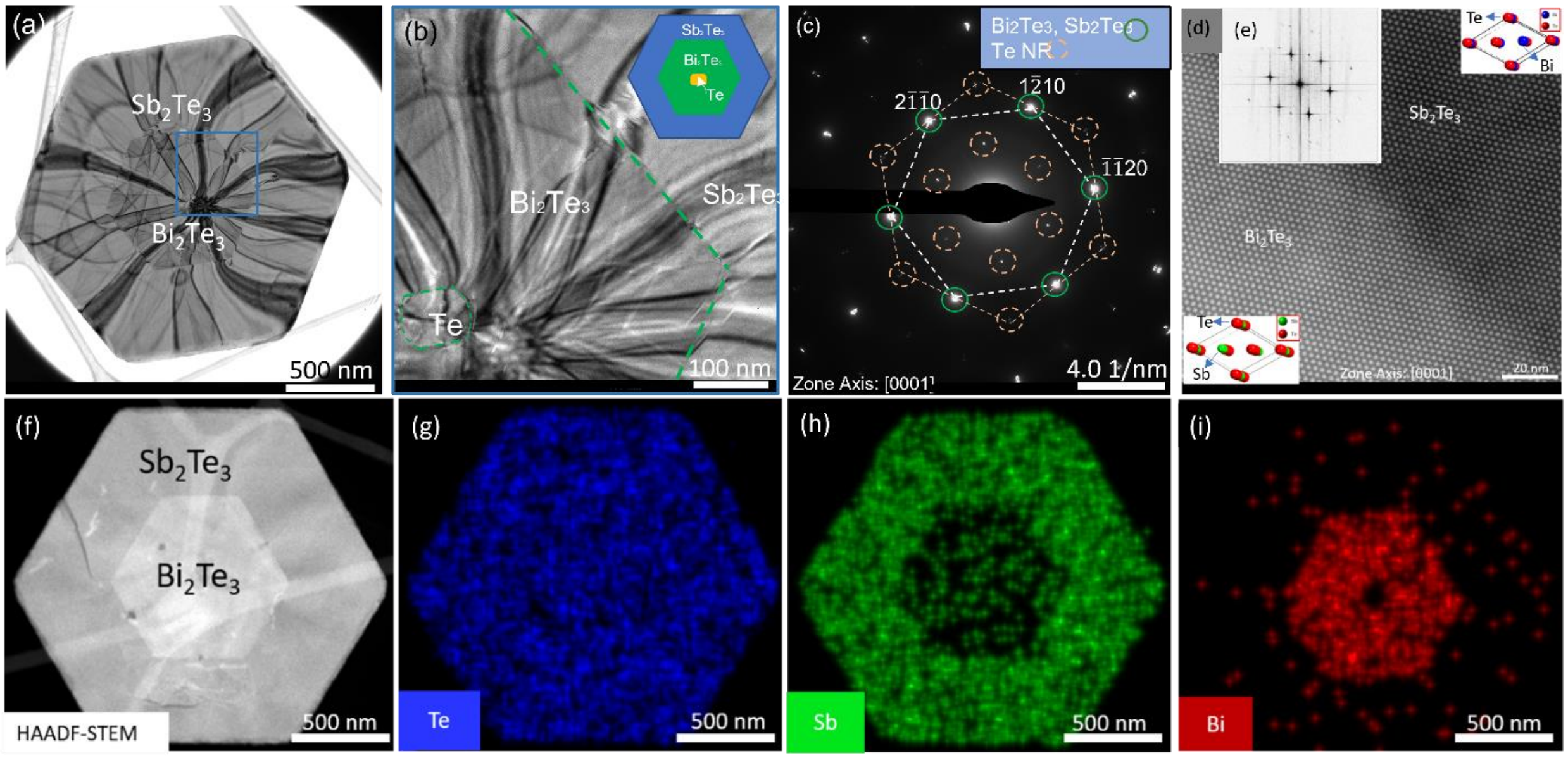

**Figure 1.** (a) TEM image of $Bi_2Te_3/Sb_2Te_3$ in-plane heterostructures displaying bending contours. (b) Magnified TEM image of the heterointerface area marked on (a) displaying the bend contours formation, originated from central Te nanorod region. (c) Electron diffraction pattern of the heterostructure (a) along the [0001] direction exhibiting diffraction from $Sb_2Te_3$ and $Bi_2Te_3$ region and Te nanorod. (d-e) HRSTEM and FFT of a heterointerface region ($Bi_2Te_3/Sb_2Te_3$) along the [0001] direction. (f) HAADF- STEM of the heterojunction flake where XEDS chemical composition maps acquired from. (g) Te XEDS chemical composition map confirming a uniform distribution of Te atoms in both $Sb_2Te_3$ and $Bi_2Te_3$ regions. (h) Sb XEDS chemical composition map confirming the composition of $Sb_2Te_3$ region. (i) Bi XEDS chemical composition map confirming the composition of $Bi_2Te_3$ region (the central Bi deficient region is due to the Te nanorod based growth process, see S2).

Presence of a near-atomically sharp heterointerface between $Bi_2Te_3$ and $Sb_2Te_3$ is confirmed using high angle annular dark field (HAADF)-STEM imaging of the heterointerface region and the corresponding fast-Fourier transform (FFT) as shown in Figure 1d and 1e. The chemical composition of the $Bi_2Te_3$-$Sb_2Te_3$ in-plane heterostructure shows the distribution of Te everywhere, while Bi and Sb are in the inner and outer regions of the heterostructure respectively (Fig.1 (f-i)). The Bi deficiency in the central region of the flake (Figure 1i) is due to the growth process of the heterostructure, starting from Te nanorods. All images presented in Figure1 have been acquired under static condition at room temperature (RT) and accelerating voltage of 80 kV. Figure 2 shows 2D $Sb_{2-x}Bi_xTe_3$ alloy heterostructures with no interface (see S5 for electron diffraction pattern acquired from the flake in Figure 2a). Bent contours are observed on multiple regions of the sample, but do not originate from the central part of the sample, contrary to the $Bi_2Te_3$-$Sb_2Te_3$ in-plane heterostructures (Figures 1(a-b)) in which the bend contours initiate from the central area inside $Bi_2Te_3$ and continue toward the outer edge of the flake. Figures 2(b-c) show high resolution scanning transmission electron microscopy (HRSTEM) images taken at two different magnifications of 2D $Sb_{2-x}Bi_xTe_3$ alloy along the [0001] zone axis with no evident trace of defects or phase segregation. The FFT pattern of the 2D $Sb_{2-x}Bi_xTe_3$ alloy from this region also shows a clear pattern of a single crystal alloy (Figure 2d). The XEDS (energy-dispersive X-ray spectroscopy) chemical analysis of this flake shows a uniform distribution of Sb, and Bi and Te at room temperature across the flake (Figure 2(e-h)). The marked region in Figure 2e should contain Te nanorod with no Bi and Sb content. To observe this compositional variation, longer acquisition time for the XEDS map is required to see if this area indeed has higher Te content than its surroundings. Figure S6 displays the two FFTs from Figure 1e and 2d superimposed in different colors. As seen from the FFTs (Figure S6), these two structures (alloy vs. heterostructure) have similar crystal structure and d-spacing.

To understand the structural dynamics, *in situ* TEM observations are conducted at elevated temperatures (see Figure S7 for low-magnification TEM image of the sample on an in-situ TEM heating device). The thermally driven structural dynamics can be seen in supporting information Movie 1, which plays the sequential frames of the $Bi_2Te_3$-$Sb_2Te_3$ heterostructure acquired between 300 °C to 390 °C at 200 kV. In addition, selected frames of the movie are shown in Figure 3(a-f) showing the sublimation process in $Bi_2Te_3$-$Sb_2Te_3$ in-plane heterostructure with the heating rate 5°C /min. The details on the experimental parameters and imaging conditions are included in materials and methods section.

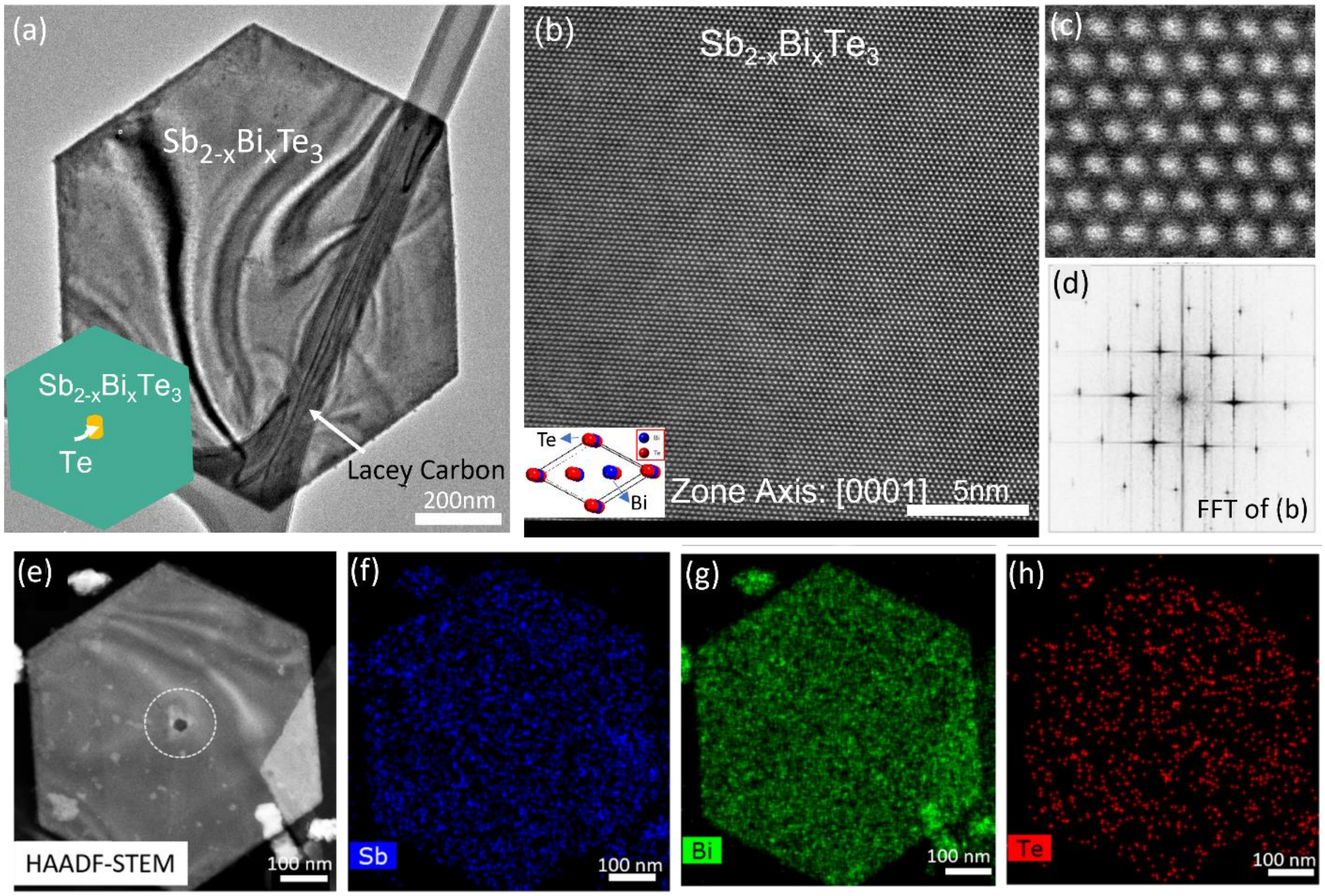

**Figure 2.** (a) TEM image of a 2D $Sb_{2-x}Bi_xTe_3$ alloy displaying bent contours originating from lacey carbon/flake interface. (b-c) HRSTEM of 2D $Sb_{2-x}Bi_xTe_3$ alloy at two different magnifications along the [0001] zone axis. (d) FFT of the HRSTEM image in (b). (e) HAADF- STEM of the alloy flake where XEDS chemical composition maps acquired from. (f) Sb XEDS chemical composition map with a uniform distribution over the flake. (g) Bi XEDS chemical composition with a uniform distribution over the flake h) Te XEDS chemical composition with a uniform distribution over the flake.

The formation of polygonal defects (nanopores) is initiated at the outer edges, $Bi_2Te_3/Sb_2Te_3$ heterointerface and the center of the $Bi_2Te_3$ with the rich Te core. Edges are one of the energetically favorable sites for sublimation due to the presence of the reactive dangling bonds and variations in bonding configurations (19). In addition, interfacial mismatch strain can induce localized structural instabilities thus making heterointerfaces, such as $Bi_2Te_3/Sb_2Te_3$, $Bi_2Te_3$/Te heterointerface, energetically favorable for the initiation of sublimation. In general, strained regions (elastic or plastic strain) can act as highly reactive areas with high thermal stress that can lead to the formation and growth of polygonal nanopores as the sublimation starts (14) (31). Overall, we speculate that under thermal stress, the high density of point defects at the Te core, interfacial mismatch strain, and chemical bonding inhomogeneities at the interfaces act as localized and reactive nanoregions that can initiate the sublimation process. In the following sections, we further link the observations on the anisotropic sublimation to the role of pre-existing point defects, various edge configurations and edge formation energies.

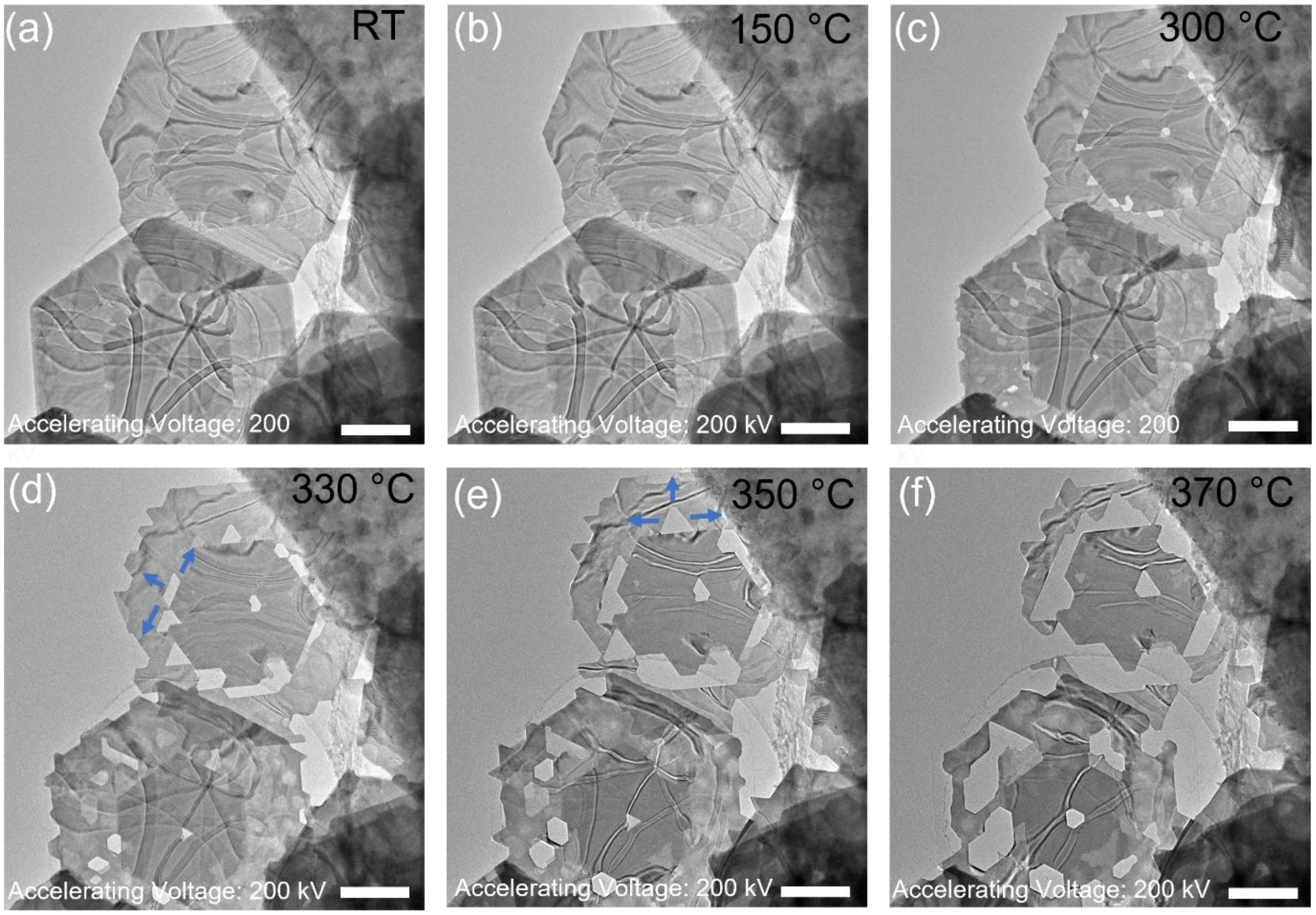

**Figure 3.** TEM images of two-dimensional $Bi_2Te_3$/$Sb_2Te_3$ heterojunction nanoplates at (a) RT (b) 150°C (c) 300 °C (d) 330°C (e) 350°C and (f) 370°C indicating the strain relaxation and sublimation of the structure through formation of preferential triangular and hexagonal nanopores forming on the heterointerface and outer edges of the structure. Blue arrows are indicating the growth direction of the polygonal nanopores (triangular and quasi-hexagonal nanopores). Scale bar is 500 nm. Images taken using accelerating voltage of 200 kV.

Triangular and quasi hexagonal nanopores are the dominant polygonal configurations observed through the sublimation. However, a transformation of nanopores can also take place as shown in Figure 3. A transformation from triangular configuration to quasi hexagonal configuration is realized for the nanopores in the Te rich core while a reverse transformation from quasi hexagonal configuration to triangular configuration occurs for the nanopores at the heterointerface. Additionally, with further increase in the temperature, these nanopore defects coalesce with the adjacent polygonal nanopores and form larger nanopores.

It is worth noting that electron beam irradiation can act as an external stimulus to trigger defect formation and growth. Therefore, to eliminate the beam effect on the formation of defects during the sublimation process, the electron beam is blanked during the *in-situ* annealing process and the specimen is only exposed when capturing an image. To confirm that the thermal annealing is the dominant external stimuli behind sublimation, a comparative in-situ annealing experiment is performed at a lower accelerating voltage of 80 kV. Thermally driven in-situ structural dynamics in $Bi_2Te_3$-$Sb_2Te_3$ in-plane heterostructure at 80 kV between 300 °C to 390 °C at 80 kV are shown in Movie 2, with selected frames of these dynamics shown in Figure 4. $Bi_2Te_3$-$Sb_2Te_3$ in-plane heterostructure at 80 kV remains stable up to 300°C before the polygonal defects start forming.

With further temperature increase, the preferential sublimation of prismatic $\{01\bar{1}0\}$ planes, such as $(1\bar{1}00)$, $(10\bar{1}0)$ and $(\bar{1}010)$, starts to occur. We observe the preferential sublimation of the edge planes of the flake to be mainly responsible for the formation of polygonal defects. With similar observation at 80 kV and 300 kV, these experiments rule out the effect of beam energy on the microstructure evolution of the flake, thus indicating the defects to be thermally driven and through the sublimation process. Figure S7c, shows a 3D schematic of the hexagonal $Bi_2Te_3$-$Sb_2Te_3$ in-plane heterostructures and $Sb_{2-x}Bi_xTe_3$ alloy displaying the crystallographic orientation of planes (zone axis: [0001]). This figure serves as a reference to identify the preferential sublimation planes during *in situ* TEM heating observations. The details on the imaging condition are included in experimental section.

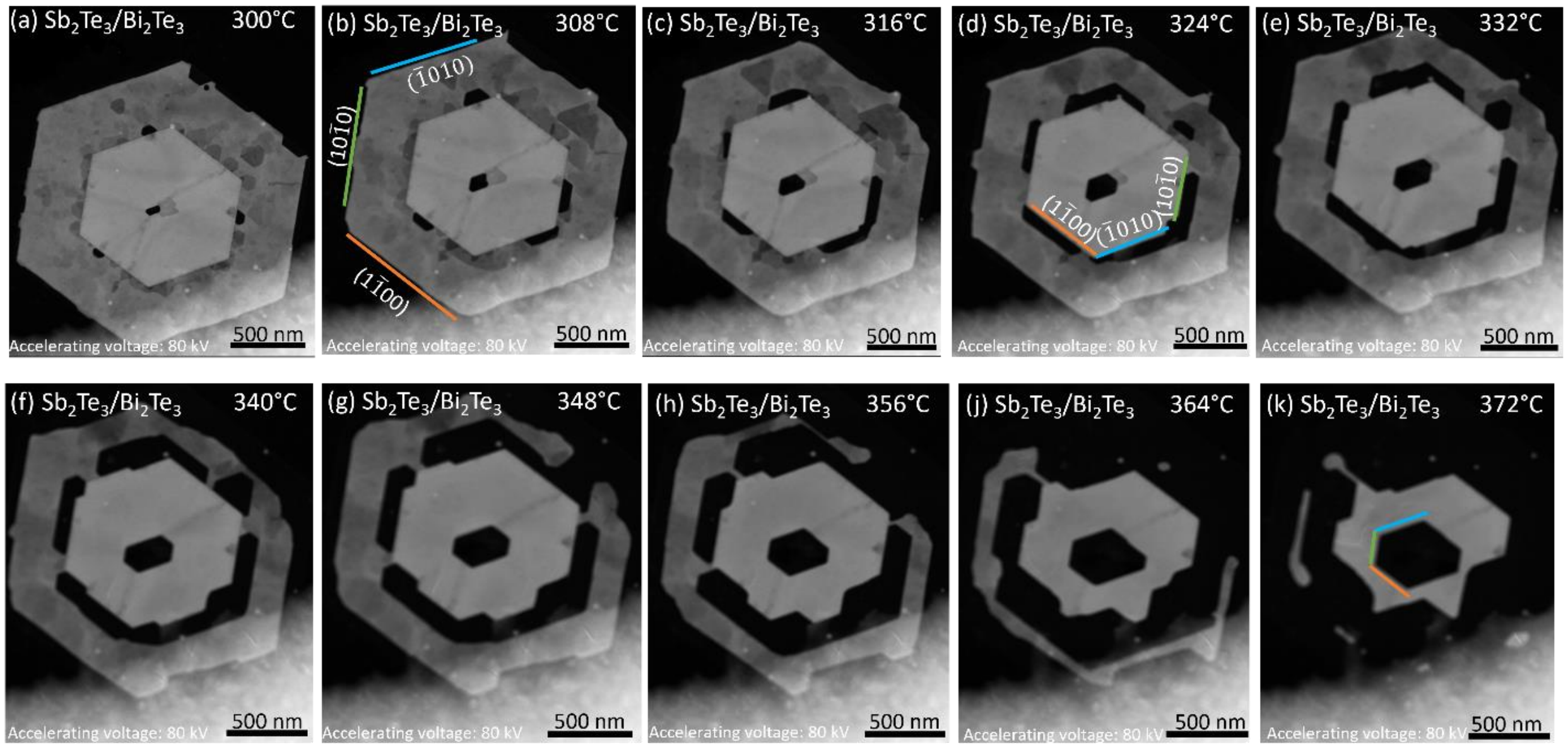

**Figure 4.** HAADF-STEM images of two-dimensional $Bi_2Te_3$/$Sb_2Te_3$ heterojunction nanoplates at (a) 300°C, (b) 308°C, (c) 316 °C, (d) 324°C, (e) 332°C, (f) 340°C, (g) 348°C, (h) 356°C, (j) 364°C and (k) 372°C, illustrating the sublimation of the heterojunction structure through formation of preferential triangular and hexagonal nanopores on the heterointerface and outer edges of the structure. Images taken using accelerating voltage of 80 kV. Image acquisition is done with an 8 °C interval and 2s exposure time/frame.

To understand how the presence of the heterointerface can alter the sublimation pathways and the sublimation starting temperature, a comparative study is performed on $Sb_{2-x}Bi_xTe_3$ ternary alloy with no heterointerface present. Thermally driven in-situ structural dynamics in $Sb_{2-x}Bi_xTe_3$ between 300 °C to 400 °C are shown in Movie 3, with selected frames of these dynamics shown here. Figure 5 presents the structural evolution upon in-situ annealing of $Sb_{2-x}Bi_xTe_3$ alloy between 310 °C and 380 °C. Selective frames of the sequential sublimation process is shown in Figure 5(a-h). Similar to the heterostructure, anisotropic and preferential sublimation of prismatic $\{01\bar{1}0\}$ planes are observed as marked on Figure 5a, b, e and h. Major differences are observed during sublimation of $Sb_{2-x}Bi_xTe_3$ alloy in comparison with the $Bi_2Te_3$-$Sb_2Te_3$ in-plane heterostructure. First, the sublimation starts at a higher temperature of around 320 °C as opposed to 300 °C for $Bi_2Te_3$-$Sb_2Te_3$ in-plane heterostructures, which can be attributed to the interfacial mismatch strain between $Bi_2Te_3$ and $Sb_2Te_3$ that can significantly impact the sublimation pathway. In contrast, in the $Sb_{2-x}Bi_xTe_3$ (nominal x=1) alloy there is no heterointerface and therefore no interfacial strain. We observe no polygonal defects to form inside the plane of the $Sb_{2-x}Bi_xTe_3$ alloy and the sublimation only takes place along the outer edges. Further in the sublimation process, a transformation from a hexagonal flake (Figure 5a) to a triangular flake (Figure 5h) is observed. The hexagonal $Sb_{2-x}Bi_xTe_3$ alloy nanoplates evolve to triangular nanoplates (retaining the low-energy facets) with 60° rotation with respect to their pristine form. The lack of internal sublimation of the alloy sample can be explained by the uniformity of the sample and reduced amount of internal local strain. Although, the sublimation drives by the chalcogenide depletion, this nanoscale study clearly indicates lattice mismatch and strain at the heterointerfaces and local structural inhomogeneities, can accelerate the sublimation and define a preferential site for anisotropic sublimation.

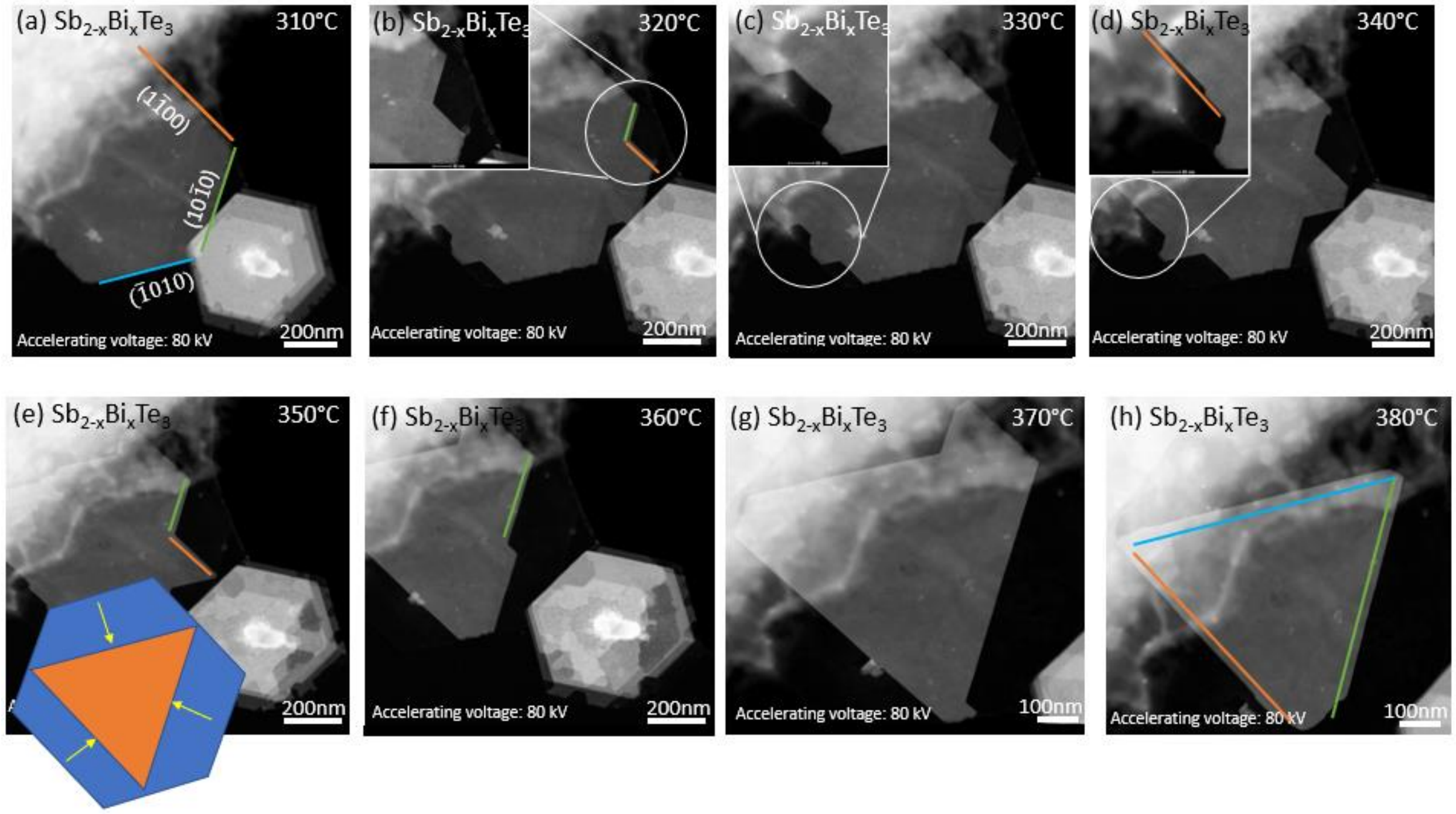

**Figure 5.** HAADF-STEM image of $Sb_{2-x}Bi_xTe_3$ alloy nanoplates at (a) 310°C, (b) 320°C, (c) 330 °C, (d) 340°C, (e) 350°C, (f) 360°C, (g) 370°C, (h) 380°C and (k)390°C. Hexagonal nanoplates evolve to triangular nanoplates with 60° rotation with respect to their pristine form through preferential sublimation of prismatic $\{01\bar{1}0\}$. Images taken using accelerating voltage of 80 kV.

Sublimation of Te and depletion of chalcogenide species at higher temperatures is anticipated to be due to the high vapor pressure of Te and weak interaction forces between $Te_1$-$Te_1$ layer in $Bi_2Te_3$ and $Sb_2Te_3$ nanostructures. Our observation of Te dissociation (chalcogenide species) is consistent with prior studies reporting on the preferential release of sulfur (S) in 2D tungsten disulfide ($WS_2$) nanodevices as the driving force for the sublimation and ultimate electrical breakdown of $WS_2$ nanodevices (14). Through sublimation process, it is expected that Te dissociates from the hexagonal nanoplates, move to the surface and by further raise in temperature, volatilizes from the surface. Sublimation of Te from the structure leaves Te dangling bonds which drives the further dissociation of Te from the structure resulting in preferential growth of polygonal defects (13) (20) (32) (33). The continuous Te loss from the structure leaves excess Bi behind (Bi bulb) as shown in XEDS chemical composition maps acquired at 360°C and 370°C (S8-S9).

Additionally, sublimation can be induced by native defects intrinsically present in the structure Density functional theory (DFT) calculations are conducted to provide an atomic insight into the structural dynamics and edge evolution during the defect aided sublimation. Possible native point defects (vacancies, antisites) as well as various point defects configurations in 2D $Bi_2Te_3$ structures are shown in Figure 6a. In addition, the associated formation energies as a function of Te chemical potential are plotted in Figure 6b. The calculations reveal that under a Bi-rich/Te-rich condition (Bi-rich refers the chemical potential of Bi ($\mu_{Bi}$) in a bulk **α**-Bi ($\mu_{Bi}^0$), i.e., $\Delta\mu_{Bi}=\mu_{Bi}-\mu_{Bi}^0=0$; Te-rich means the chemical potential of Te ($\mu_{Te}$) in a bulk **α**-Te ($\mu_{Te}^0$), i.e., $\Delta\mu_{Te}=\mu_{Te}-\mu_{Te}^0=0$) antisite defects are the dominant native defects in $Bi_2Te_3$ structure. This is consistent with the previous findings by Scanlon et. al (34) showing that the antisite defects play a key role in controlling the bulk conductivity of topological insulators. The calculations further reveal, in 2D $Bi_2Te_3$ nanoplates under Te-poor condition, $Bi_{Te1}$ antisite defect where Te is replaced by Bi possesses the lowest formation energy among all native defects while under Te-rich condition, $Te_{Bi}$ antisite defect in which an excess Te occupies the Bi site is the most energetically favorable point defect. Note that the vacancy defect $V_{Te1}$ is the most energetically favorable when $\Delta\mu_{Te}$ is in the range of –0.28 - –0.16 eV. This is different from that for bulk $Bi_2Te_3$ where the formation energies for all three types of vacancies ($V_{Bi}$, $V_{Te1}$ and $V_{Te2}$) are found to be much higher than that for antisites in both Te-rich and Bi-rich conditions (35).

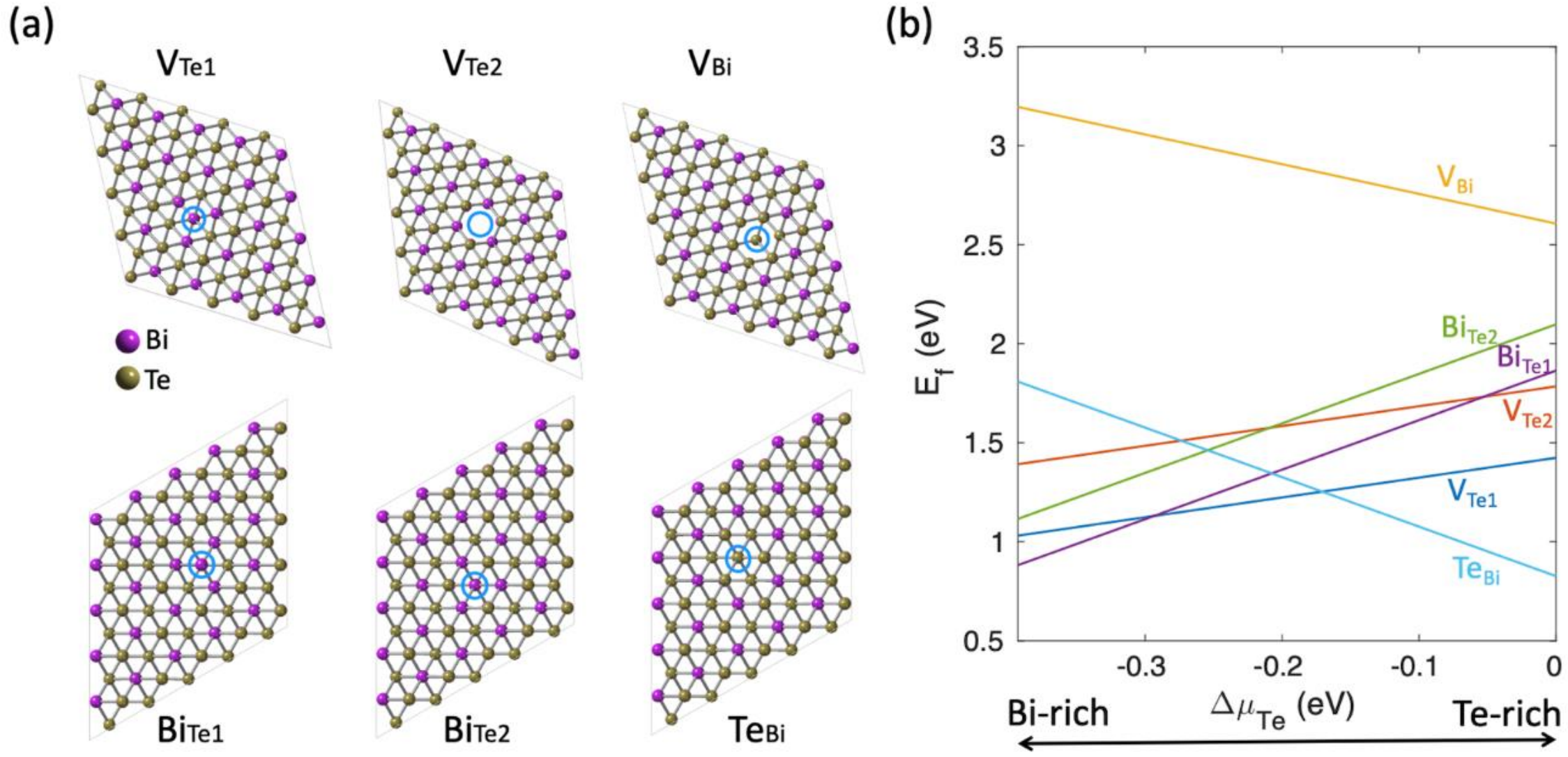

**Figure 6.** Point defects in 2D (1QL) $Bi_2Te_3$ investigated at the DFT level. (a) Native point defects configuration including vacancies and antisite defects. (b) The formation energy of vacancy and antisites defects in 2D $Bi_2Te_3$ as a function of chemical potential of Te. $Te_{Bi}$ antisite are dominant defects under the Te rich/Bi poor condition.

Due to the abundance of Te through the synthesis procedure, epitaxial nucleation, and growth of $Bi_2Te_3$ on the surface of Te nanorods, the represented in-plane $Bi_2Te_3$-$Sb_2Te_3$ heterostructure and $Sb_{2-x}Bi_xTe_3$ alloy in our study are representing a Te rich scenario, therefore $Te_{Bi}$ antisite defects are the most dominant native defects present in these structures at room temperature (RT). Since an excess Te atom weakly interact with two adjacent Te host atoms in the lattice, as compared to the Bi-Te covalent bond, we expect that the presence of $Te_{Bi}$ antisite promotes the Te-sublimation. It is also likely that this antisite would be a perfect nucleation site for in-plane heterostructure growth.

Various atom-detachment pathways can result in the formation of different type of edge structures during sublimation. The edge configurations are mediated by the chemical potential difference between Bi and Te in $Bi_2Te_3$ quintuple (QL). For $Bi_2Te_3$ QL, eight edge models are generated and selectively presented in Figure S10-11: zigzag (ZZ) edges with Te and/or Bi terminations (ZZ-4Te, ZZ-3Te1Bi, ZZ-4Bi, and ZZ-6Te-sat (fully saturated)), and armchair edges (AC-0 (stoichiometric), AC-0-sat (fully saturated), AC-1, AC-2 and AC-3). Each model is allowed to relax to their ground state. Two representatives of edge configurations (ZZ-4Te and AC-0) are shown in Figure 7a Figure 7b presents the edge formation energies as a function of the excess of chemical potential of Te ($\Delta\mu_{Te}$). The formation energy for stoichiometric armchair edge (AC-0) is a constant number with respect to the $\Delta u_{Te}$ and it is also thermodynamically most stable in allowed Te chemical potential, indicating that the morphology of $Bi_2Te_3$ should be dominated by the AC-0 edge during the growth. The next energetically favorable edge is the zigzag ZZ-4Te in Te-rich condition, and armchair edge AC-2 in Bi-rich condition. The formation energy for ZZ-4Bi edge is always the highest in allowed Te/Bi conditions, indicating that the ZZ-4Bi is less likely to be formed during growth; while the formation of other edges, like AC-3, AC-0-sat, ZZ-6Te-sat, AC-1 and AC-2 are highly depending on the Te-rich or Bi-rich condition.

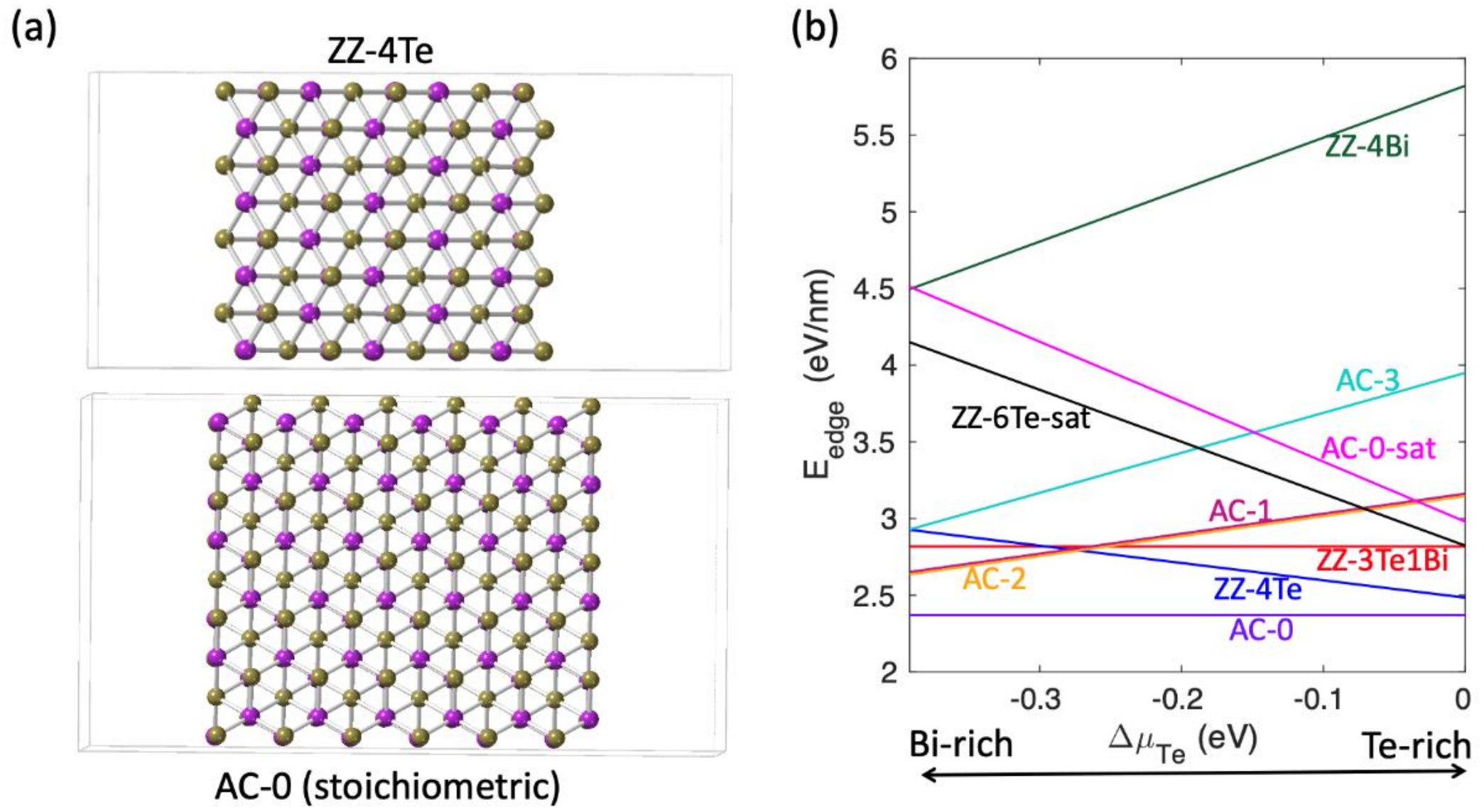

**Figure 7.** (a) Representatives for zigzag (ZZ-4Te) and armchair (AC-0) edges for 2D $Bi_2Te_3$ QL. (b) Formation energies for different types of zigzag and armchair edges with structures shown in Figure S10-11.

Slow-growing edges are thermodynamically the most stable ones and therefore the most energetically favorable ones during growth and dominating the growth morphology. However, during sublimation and etching, it is expected that the fast-growing edges with higher formation energies govern both the sublimation process and the nanopores morphology. The hexagonal morphology of the flake as the equilibrium shapes at room temperature for both the alloy and heterostructure can further confirm that the armchair edges (AC-0) are the most stable edges at RT. However, it is expected that at higher temperatures both the zigzag edges (ZZ-4Bi, ZZ-6Te-sat) and armchair edges (AC-3, AC-0-sat) could be the more stable edges of choice. This is consistent with the sublimation observations that is shown in Figure 5, displaying a hexagonal to triangular transformation for the alloy structure at higher temperatures.

## Conclusion

We have reported the direct observation of sublimation dynamics pathways in a $Bi_2Te_3$-$Sb_2Te_3$ in-plane heterostructure and its ternary alloy counterpart via *in situ* TEM at elevated temperatures. The microstructural evolution consisted of formation, growth and coalescence of thermally induces nanopores and ultimately preferential sublimation initiating at reactive regions with structural heterogeneities. The preferential sublimation sites are center, heterointerface and outer edges for the heterostructure and only outer edges for the alloy counterpart. Excessive Te in the center of the structure and high density of defects on the heterointerface are anticipated to be the main driving force for preferential depletion of chalcogenide (Te) from these reactive sites. Triangular and quasi hexagonal configurations are observed to be the dominant configurations, meanwhile transformation between different configurations is also reported. Furthermore, density functional theory calculations unravel that antisite defects ($Te_{Bi}$), in which excess Te occupies the Bi site in the lattice, possess the lowest formation energy therefore are expected to be the dominant native defects in the structure and play a key rule on the sublimation process. In addition, based on the calculated edge formation energies for various edge configurations, it is expected that at higher temperatures both the zigzag edges (ZZ-4Bi, ZZ-6Te-sat) and armchair edges (AC-3, AC-0-sat) be the more stable edges of choice. An in-depth understanding of the nanoscale dynamics is the key for making tunable layered metal chalcogenides with extended functionalities from optoelectronic devices to efficient catalyst for energy conversion. Furthermore, thermal induced defects as physical inclusions can provide pathways offering further tunability in these systems.

## Materials and Methods

**Preparation of $Bi_2Te_3$-$Sb_2Te_3$ in-plane heterostructure and $Sb_{2-x}Bi_xTe_3$ alloy.** Hexagonal 2D $Bi_2Te_3$ nanoplates are synthesized using a solution phase solvothermal process and based on Te-seeded growth. Tellurium oxide ($TeO_2$) and bismuth oxide ($Bi_2O_3$) used as the major precursors for this synthesis matching the stoichiometry ratio at 180 $^oC$. A dispersed solution is acquired by mixing and stirring the following components, based on the reported ratio and concentration at 100 $^oC$: 0.48 g of polyvinylpyrrolidone, 30 ml ethylene glycol, 15 mM sodium hydroxide (NaOH), 1.7 mM tellurium oxide ($TeO_2$) and 0.6 mM bismuth oxide ($Bi_2O_3$). The solution is further annealed at 180 $^oC$ for 15 h. A schematic of $Bi_2Te_3$ nanosheets nucleation and growth from the surface of Te nanorods is included in Figure S2. The material is then washed thoroughly via Deionized (DI) water following with an immediate freeze-drying procedure to prevent oxidation. The $Bi_2Te_3$ nanosheets kept under inert atmosphere of argon inside the glovebox. For $Sb_{2-x}Bi_xTe_3$ alloy synthesis, antimony oxide ($Sb_2O_3$) precursor is simultaneously mixed with $Bi_2O_3$ precursor with a 1:1 ratio. However, for $Bi_2Te_3$-$Sb_2Te_3$ heterostructure synthesis, $Bi_2Te_3$ nanosheets are used as seeds to grow a $Sb_2Te_3$ on the outer edges of $Bi_2Te_3$ , during which $Sb_2O_3$ and $TeO_2$ as precursors are used for the growth of $Sb_2Te_3$ (6) (24).

**Characterization (instrumentation and data acquisition).** TEM samples are prepared using solvent assisted exfoliation (a 1:1 mixture of Isopropyl Alcohol (IPA) and DI water) following with 30-45 minutes of sonication to acquire a clear solution. A drop of the clear solution is dropped over a Fusion in-situ TEM heating device and is dried at room temperature. Talos F200X scanning/transmission electron microscope (STEM) equipped with a heating stage and a super XEDS detector at accelerating voltage of 80 kV is used to acquire cross-sectional STEM images and XEDS maps. Microstructural evolution of in-plane $Bi_2Te_3$-$Sb_2Te_3$ heterostructure and $Sb_{2-x}Bi_xTe_3$ alloy are studied using high angle annular dark field (HAADF)-STEM and TEM imaging techniques combined with electron diffraction pattern (EDP). Chemical compositions of the nanoplates are further evaluated using XEDS. To minimize the electron beam irradiation and prevent beam-induced defects mainly knock-on damage in this case, in-situ experiments is mainly carried out at 80 kV. However, a comparative in-situ TEM study is carried out at 200 kV for the $Bi_2Te_3$-$Sb_2Te_3$ heterostructure. During annealing, the electron beam is blanked to ensure that structural dynamics are dominantly a function of thermal annealing and minimize the electron beam irradiation effect. The in-situ annealing experiments are performed under vacuum with no chemical agent. Experimental parameters such as heating rate (5°C /min) and exposure time (2s) per frame kept consistent with the in-situ annealing experiments at 200 kV.

**Density Functional Theory (DFT) Simulation.**

All calculations for the energies of native defects and edges for 2D $Bi_2Te_3$ QL are carried out in the framework of density functional theory (DFT) (36) implemented in the Vienna Ab initio Simulation Package (VASP) (37) (38) using the projector augmented waves (PAW) (39) method and the Perdew–Burke–Ernzerhof (PBE) (40) generalized gradient approximation (GGA) (40) exchange-correlation energy functional. A 5×5×1 supercell of 2D $Bi_2Te_3$ QL structure is generated and then used for created single defects including vacancies and antisites. All structures are fully relaxed with a force criterion of 0.01 eV/Å/atom. The energy cutoff for the plane-wave basis is set to be 450 eV and the Monkhorst–Pack k-point is 4×4×1 for calculating free energies of the pristane and defect structures. The formation energy for the defect structure is obtained using equation (1), where $E_{defect}$ and $E_{pristine}$ are the energy of one supercell (1 QL) with and without a defect respectively, $\Delta n_i$ and $\mu_i$ are the change in the number of atoms and chemical potential of element i, respectively. The chemical potentials of elements Bi ($\mu_{Bi}$) and Te ($\mu_{Te}$) can be written as $\mu_{Bi}= \mu_{Bi}{}^0+\Delta\mu_{Bi}$ and $\mu_{Te}= \mu_{Te}{}^0+\Delta\mu_{Te}$, where $\mu_{Bi}{}^0$ and $\mu_{Te}{}^0$ are the refence chemical potentials of bulk Bi and bulk Te, respectively. Then Equation (1) can be rewritten to equation (2) as a function of $\Delta\mu_{Te}$, where $\Delta G^0$ is the calculated Gibbs free energy for forming one formula unit of $Bi_2Te_3$ expressed in Equation (3) with $\mu_{Bi_2Te_3}$ being the chemical potential of one formula unit of $Bi_2Te_3$. Ribbons with different zigzag or armchair edges are generated with 15 Å separation between the edge and its image, and then optimized with energy cutoff of 400 eV and Kpoints of 1×6×1 and 7×1×1 for armchair and zigzag edges, respectively. The formation energy of edges is calculated using equation (4), where L is the length of the edge, $E^{ribbon}$ is the total energy of the ribbon, $n_i$ is the number of atoms of element i. It could be rewritten as a function of $\Delta\mu_{Te}$ in Equation (5).

$$E^f = E^{defect} - E^{pristine} - \sum_i \Delta n_i \, \mu_i \quad (1)$$

$$E^f = E^{defect} - E^{pristine} - \sum_i \Delta n_i \, \mu_i{}^0 - \frac{1}{2}\Delta G^0 \Delta n_{Bi} - \left(\Delta n_{Te} - \frac{3}{2}\Delta n_{Bi}\right)\Delta\mu_{Te} \quad (2)$$

$$\Delta G^0 = \mu_{Bi_2Te_3} - 2\mu^0_{Bi} - 3\mu^0_{Te} = 2\Delta\mu_{Bi} + 3\Delta\mu_{Te} \quad (3)$$

$$E_{edge} = \frac{1}{2L}\left(E^{ribbon} - \sum_{i} n_i \ \mu_i\right) \quad (4)$$

$$E_{edge} = \frac{1}{2L}\left(E^{ribbon} - \frac{1}{2}n_{Bi}\ \boldsymbol{\mu_{Bi_2Te_3}} + (\frac{\mathbf{3}}{\mathbf{2}}n_{Bi} - n_{Te})(\boldsymbol{\mu^0_{Te}} + \boldsymbol{\Delta\mu_{Te}})\right) \quad (5)$$

## ACKNOWLEDGMENT

P.M and N.A acknowledge Penn State MRSEC, Center for Nanoscale Science, under the award NSF DMR-1420620, T.S. acknowledge Swedish Research Council (Grant No. 2015-06462 and 2015-00520). P.M.A. acknowledges support from the Air Force Office of Scientific Research under award number FA9550-18-1-0072. ACTvD, NN and TW acknowledge funding from the acknowledge the funding from the National Science Foundation 2D Crystal Consortium Materials Innovation Platform (NSF 2DCC-MIP) under cooperative agreement DMR-1539916. Computations for this research were performed on the PSU's Institute for Cyber Science Advanced Cyber Infrastructure (ICS-ACI).

## ABBREVIATIONS

*In situ* TEM, *In Situ* Transmission Electron Microscopy; STEM, Scanning Transmission Electron Microscopy; DFT, Density Functional Theory; QL, Quintuple; HAADF, High Angle Annular Dark Field; XEDS, X-ray Energy Dispersive Spectroscopy; AC, Armchair; ZZ, Zigzag.

## ASSOCIATED CONTENT

### Supporting Information

The supporting information is available free of charge online. The supplementary information includes Figures S1−S12. In addition, movie 1-3 are also available as web enhanced objects free of charge.

## AUTHOR INFORMATION

### Corresponding Authors

*Nasim Alem (email: nua10@psu.edu) and Parivash Moradifar (email: pmoradi@stanford.edu).

### Author Contributions

P.M. designed the study in consultation with N.A. P.M. carried out the *in situ* TEM experiments and data analysis. T.W., A.V.D. and N.N carried out the DFT calculations. T.S. and P.A. carried out the nanoplates synthesis work. The manuscript was drafted by P.M. and edited by all co-authors.

# *Supporting Information*

## Table of Contents

**Figure S8.** XEDS chemical composition maps at (a) 360 C and (b) 370 C from two flakes 2D $Bi_2Te_3$-$Sb_2Te_3$ in-plane heterostructure, revealing the depletion of Te (chalcogenide species) during sublimation.

**Figure S9.** Excessive depletion of Sb and Te resulting in formation of a Bi bulb. (a) HAADF-STEM images of $Sb_{2-x}Bi_xTe_3$ alloy at ~380°C, (b) HAADF-STEM images of $Sb_{2-x}Bi_xTe_3$ alloy at ~300 °C, (c) EDS map of Bi at ~380°C, (d) EDS map of Te at ~380°C, e) EDS map of Sb at ~380°C.

**Figure S10.** Configuration of various zigzag (ZZ) edges in $Bi_2Te_3$ investigated in the DFT framework.

**Figure S11.** Configuration of various armchair (AC) edges in $Bi_2Te_3$ investigated in the DFT framework.

**Figure S12. (**a) Orientation of AC edges, (b) Orientation of ZZ-Bi edges, (c) Orientation of ZZ-Te edges, (d) Orientation of AC/ZZ edges, (e) Summary table of edge type vs. orientations.

**Table 1.** Lattice parameters for $Bi_2Te_3$ and $Sb_2Te_3$

**Formation of solvothermally synthesized hexagonal platelets**.

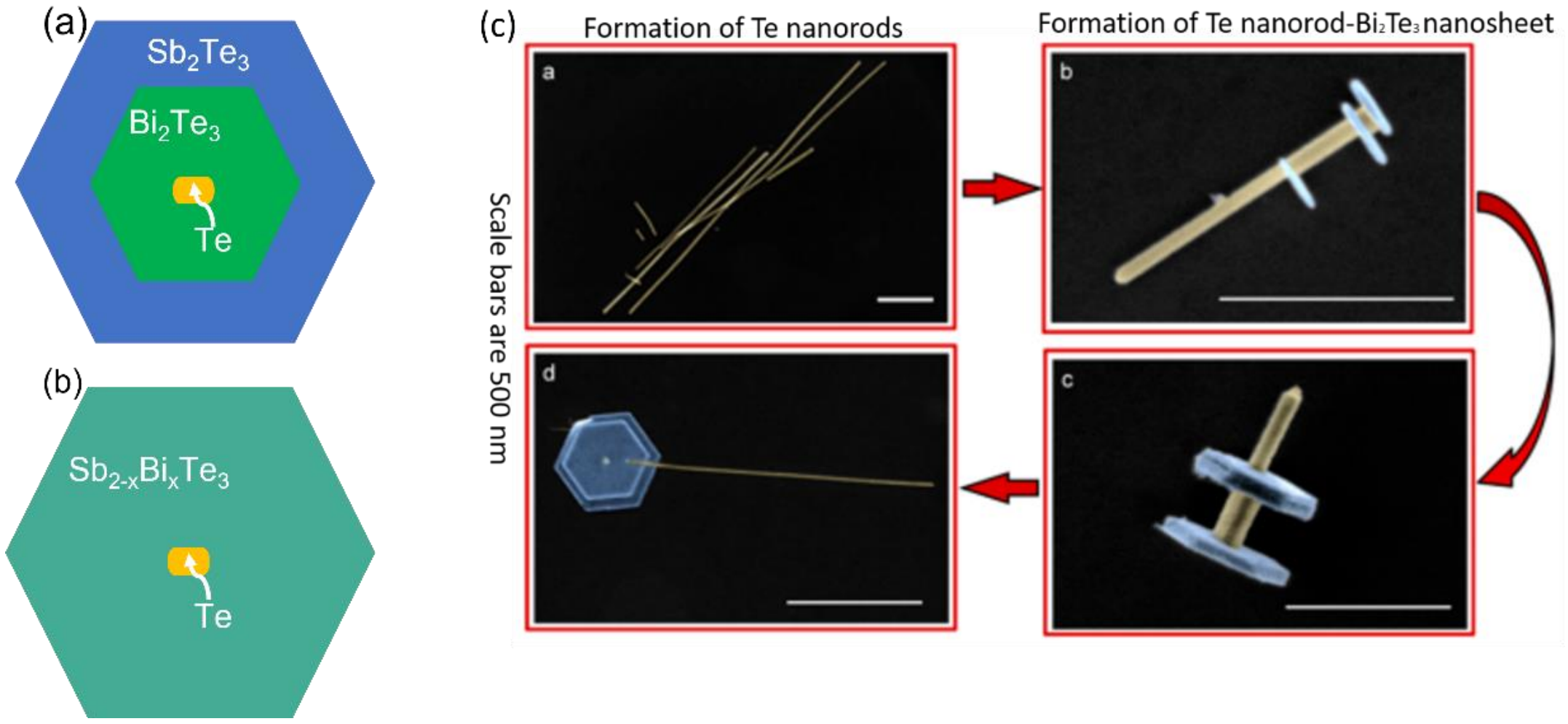

**Figure S1.** Simple Schematic of (a) $Bi_2Te_3$/$Sb_2Te_3$ in-plane heterostructures and (b) $Sb_{2-x}Bi_xTe_3$ alloy (c) Solvothermal synthesis procedure of $Bi_2Te_3$/$Sb_2Te_3$ in-plane heterostructures and $Sb_{2-x}Bi_xTe_3$ alloy based on Te-seeded growth and with controlling nucleation driving force in two dimensions and suppressing in the third dimension.

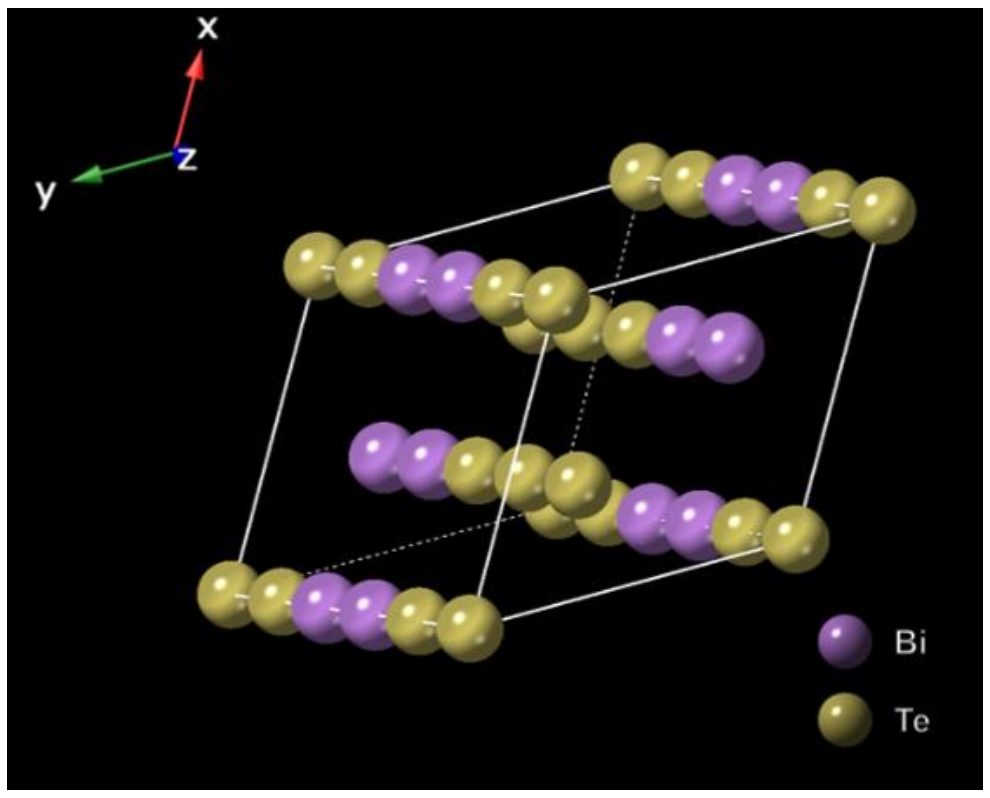

**Figure S2.** Crystal structure of $Bi_2Te_3$ and $Sb_2Te_3$. Rhombohedral crystal structure of $Bi_2Te_3$ and $Sb_2Te_3$ with space group: R$\bar{3}$m.

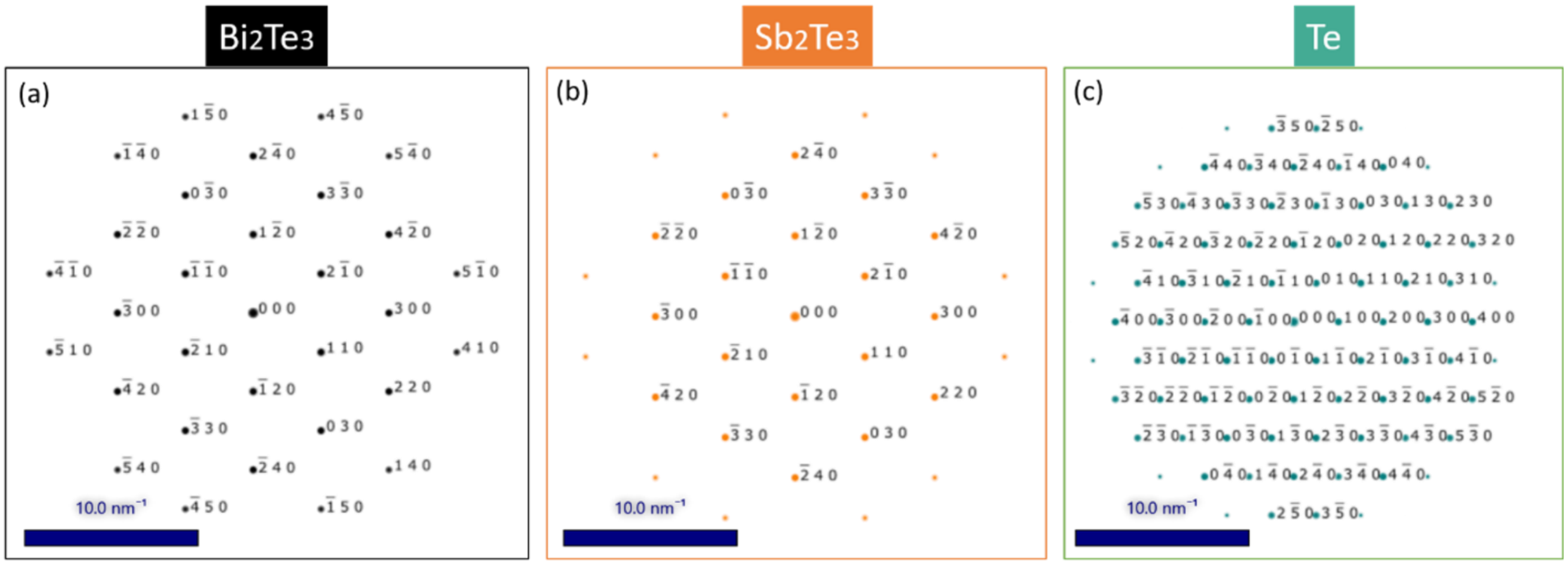

**Figure S3.** Simulated electron diffraction patterns (EDPs) (a) simulated electron diffraction pattern for $Bi_2Te_3$ along zone axis [001], (b) simulated electron diffraction pattern for $Sb_2Te_3$ along zone axis [001], (c) simulated electron diffraction pattern for Te nanorod along zone axis [001].

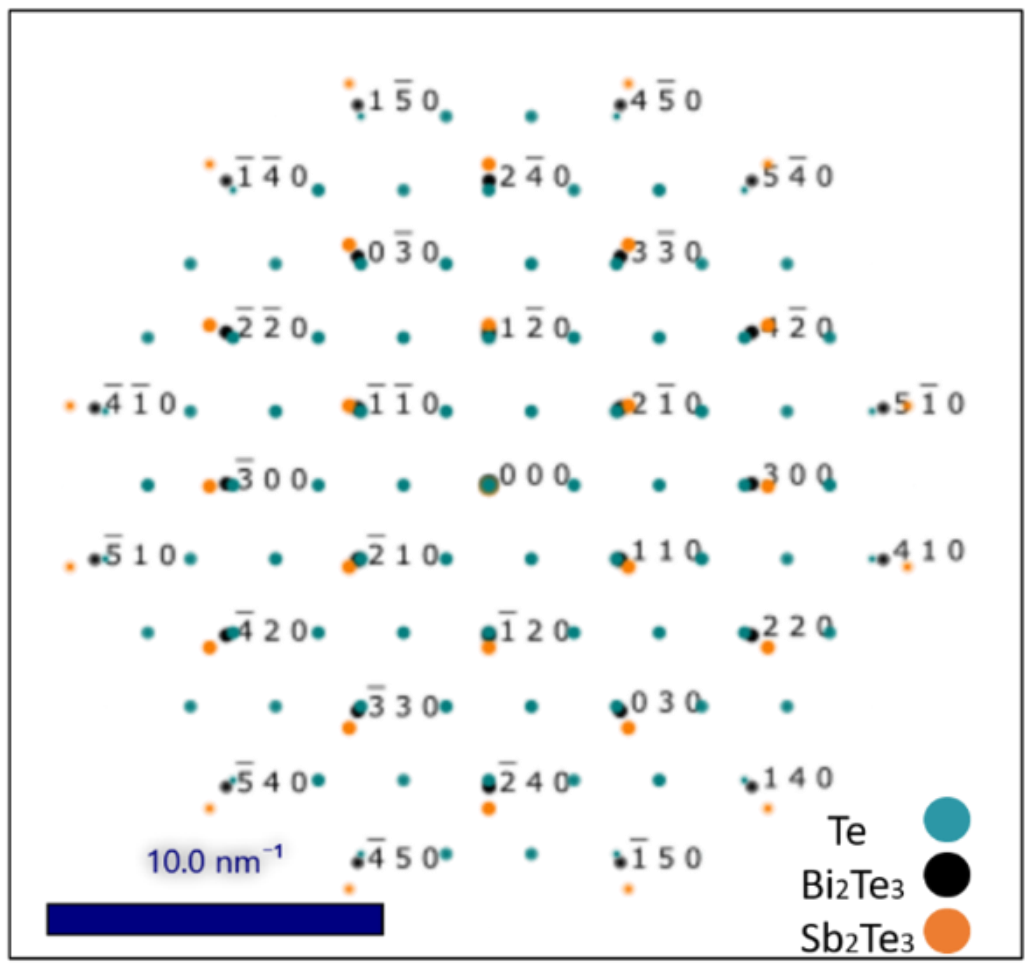

**Figure S4.** Superimposed electron diffraction patterns (EDPs) for $Bi_2Te_3$, $Sb_2Te_3$ and Te nanorod. Diffraction spots for $Bi_2Te_3$ and $Sb_2Te_3$ are overlapping and Te cause extra diffraction spots with the same symmetry.

The lattice parameters for $Bi_2Te_3$ and $Sb_2Te_3$ are $a_{Bi2Te3}$= $b_{Bi2Te3}$= 4.384 Å, $c_{Bi2Te3}$= 30.487 Å, $a_{Sb2Te3}$= $b_{Sb2Te3}$= 4.264 Å, $c_{Sb2Te3}$ = 30.458 Å (24) and are listed in Table 1. The lattice mismatch between $Bi_2Te_3$ and $Sb_2Te_3$ is about 2.5%.

**Table 1.** Lattice Parameters for $Bi_2Te_3$ and $Sb_2Te_3$

| Lattice Parameters | $Bi_2Te_3$ | $Sb_2Te_3$ |
|---|---|---|
| a | 4.384 Å | 4.264 Å |
| b | 4.384 Å | 4.264 Å |
| c | 30.487 Å | 30.458 Å |

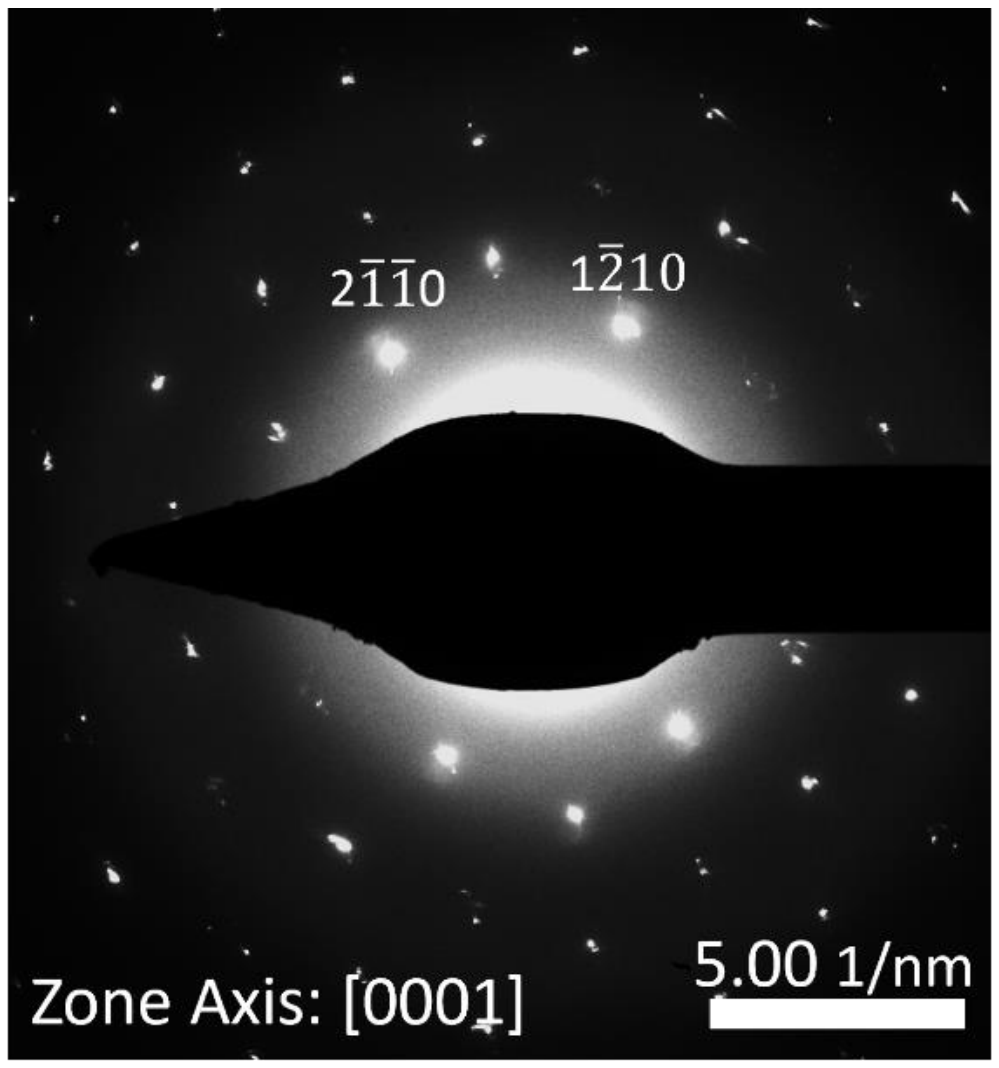

**Figure S5.** Electron diffraction pattern of $Sb_{2-x}Bi_xTe_3$ alloy and Te nanorod. Diffraction spots Te cause extra diffraction spots with the same symmetry.

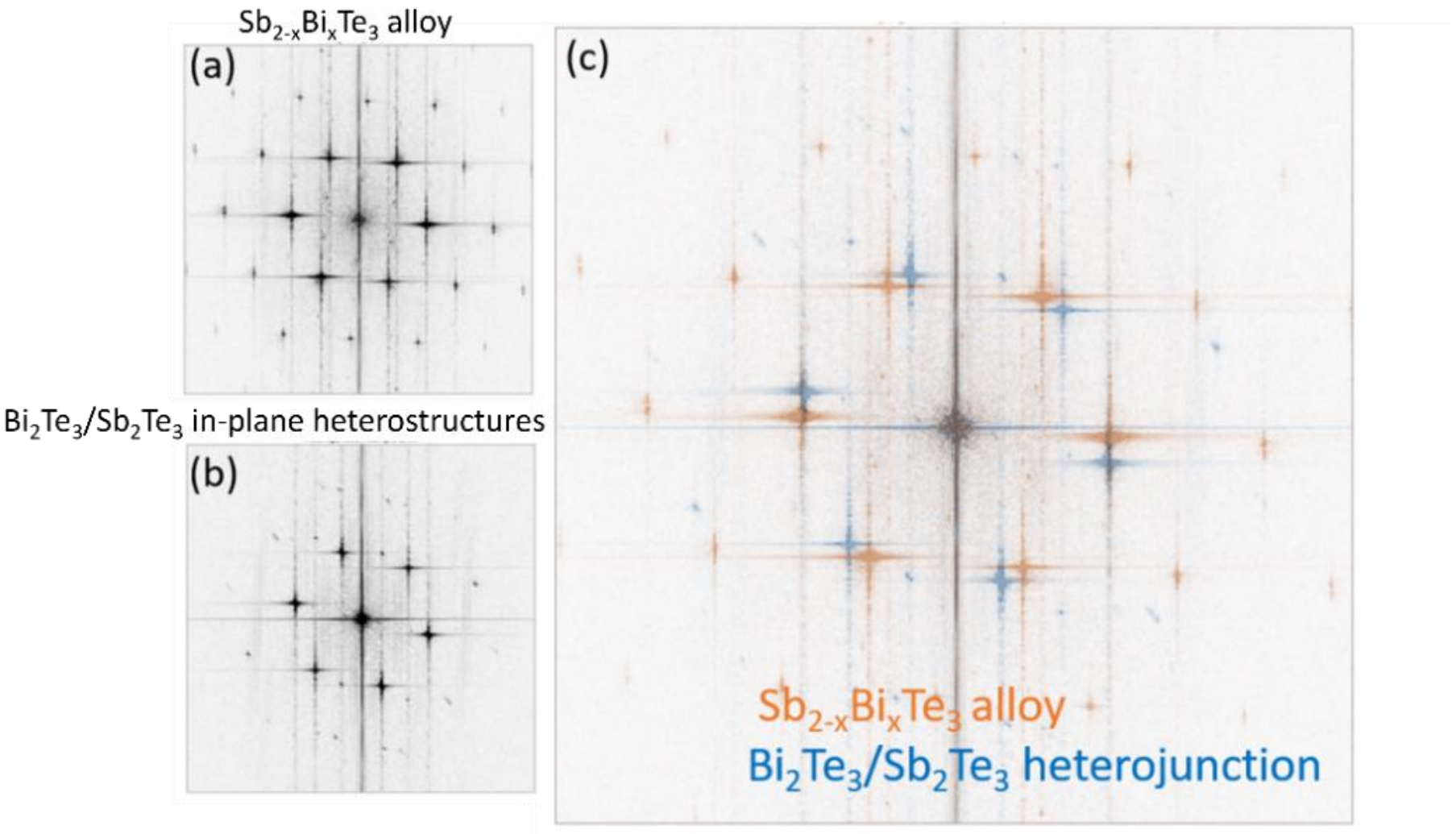

**Figure S6.** Compared FFTs of 2D $Sb_{2-x}Bi_xTe_3$ alloy and $Bi_2Te_3/Sb_2Te_3$ in-plane heterostructures (a) FFT of the 2D $Sb_{2-x}Bi_xTe_3$ alloy(2d), (b) FFT of the 2D $Bi_2Te_3$-$Sb_2Te_3$ in-plane heterostructure (1e), (c) superimposed FFTs (a) and (b) displaying similar structure with slight rotation.

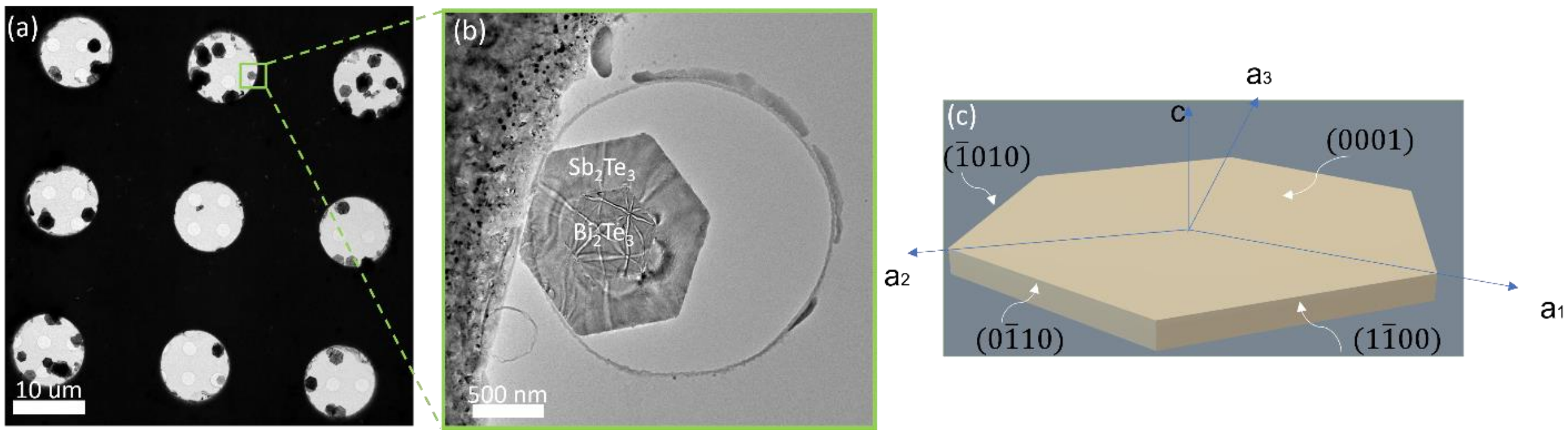

**Figure S7.** Sample preparation of $Bi_2Te_3$-$Sb_2Te_3$ in-plane heterostructures for in-situ TEM annealing experiment, (a) low-mag HAADF-STEM of $Bi_2Te_3$-$Sb_2Te_3$ heterojunction flakes dispersed on the Protochips-heating device, (b) $Bi_2Te_3$-$Sb_2Te_3$flake over a hole covered with ultra-thin and electron-transparent amorphous carbon layer, (c) 3D schematic of the $Bi_2Te_3$-$Sb_2Te_3$ in-plane heterostructures and $Sb_{2-x}Bi_xTe_3$ alloy. 3D schematic used to identify the preferential sublimation planes during in-situ TEM annealing.

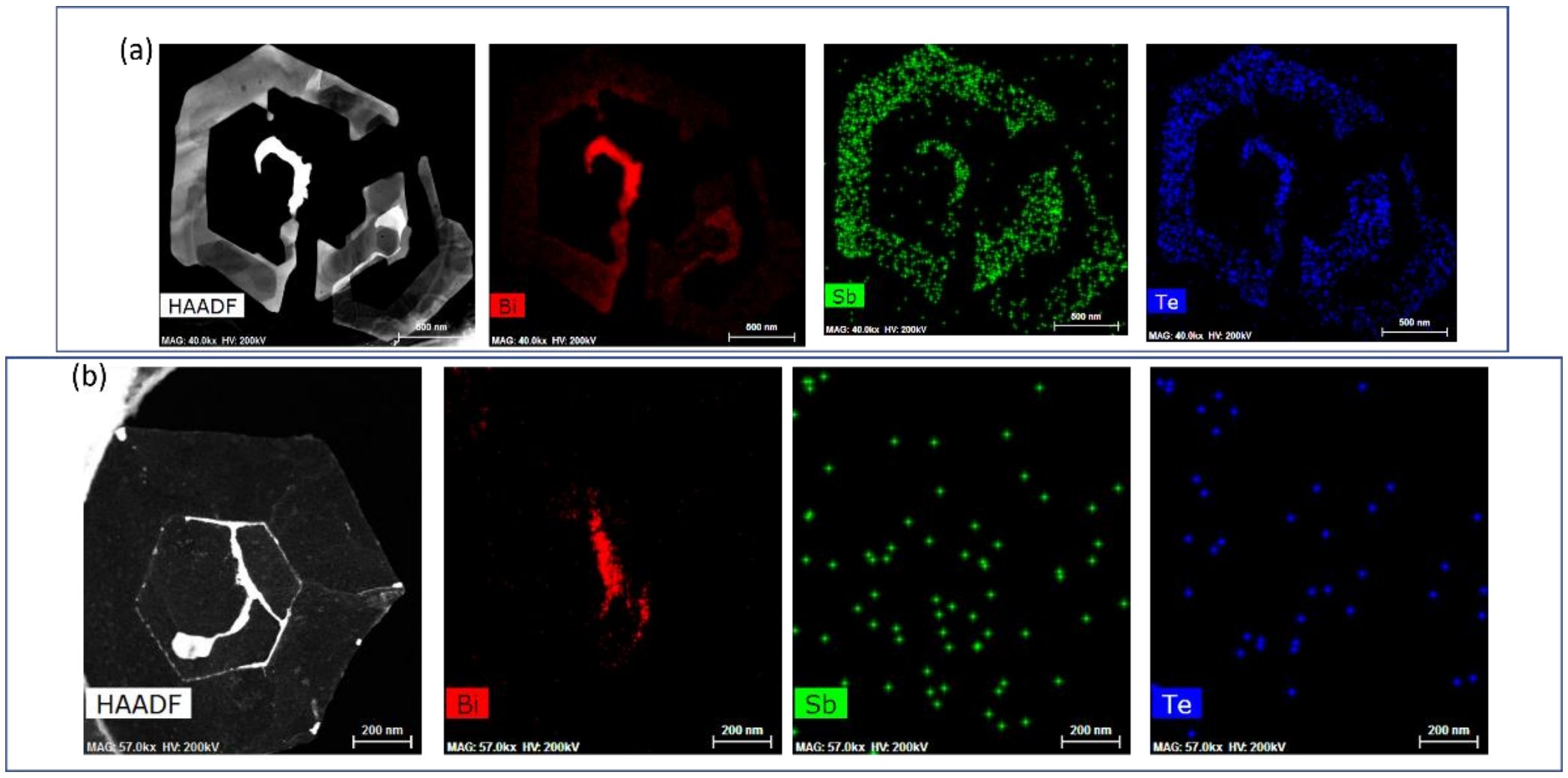

**Figure S8.** XEDS chemical composition maps at (a) 360 C and (b) 370 C from two flakes 2D $Bi_2Te_3$-$Sb_2Te_3$ in-plane heterostructure, revealing the depletion of Te (chalcogenide species) during sublimation.

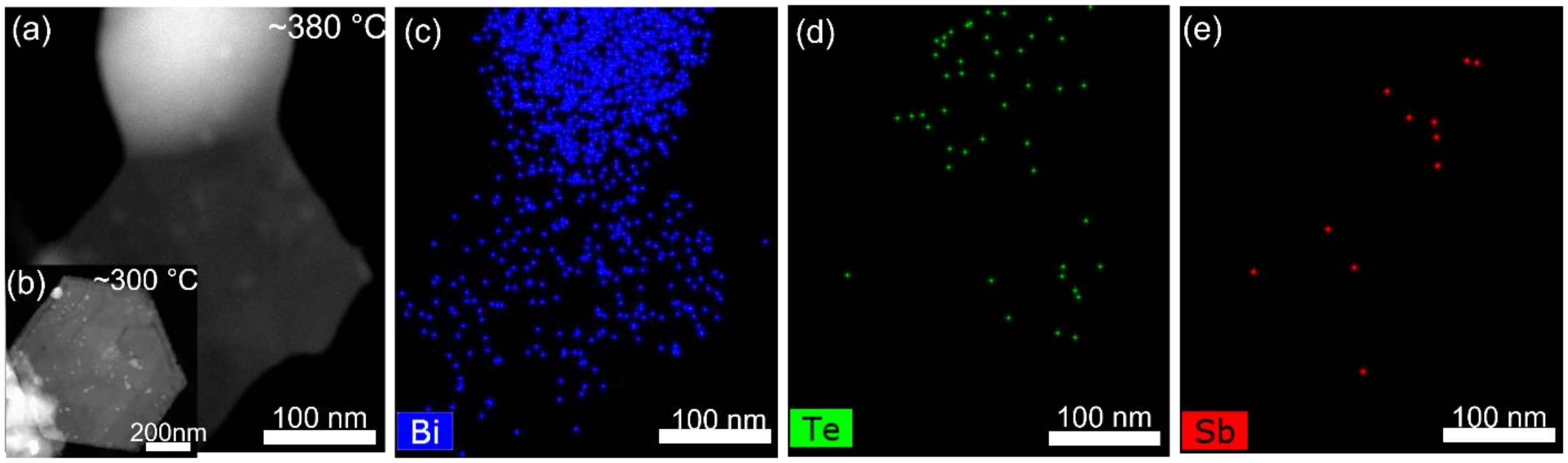

**Figure S9.** Excessive depletion of Sb and Te resulting in formation of a Bi bulb, (a) HAADF-STEM images of $Sb_{2-x}Bi_xTe_3$ alloy at ~380°C, (b) HAADF-STEM images of $Sb_{2-x}Bi_xTe_3$ alloy at ~300 °C, (c) EDS map of Bi at ~380°C, (d) EDS map of Te at ~380°C, (e) EDS map of Sb at ~380°C.

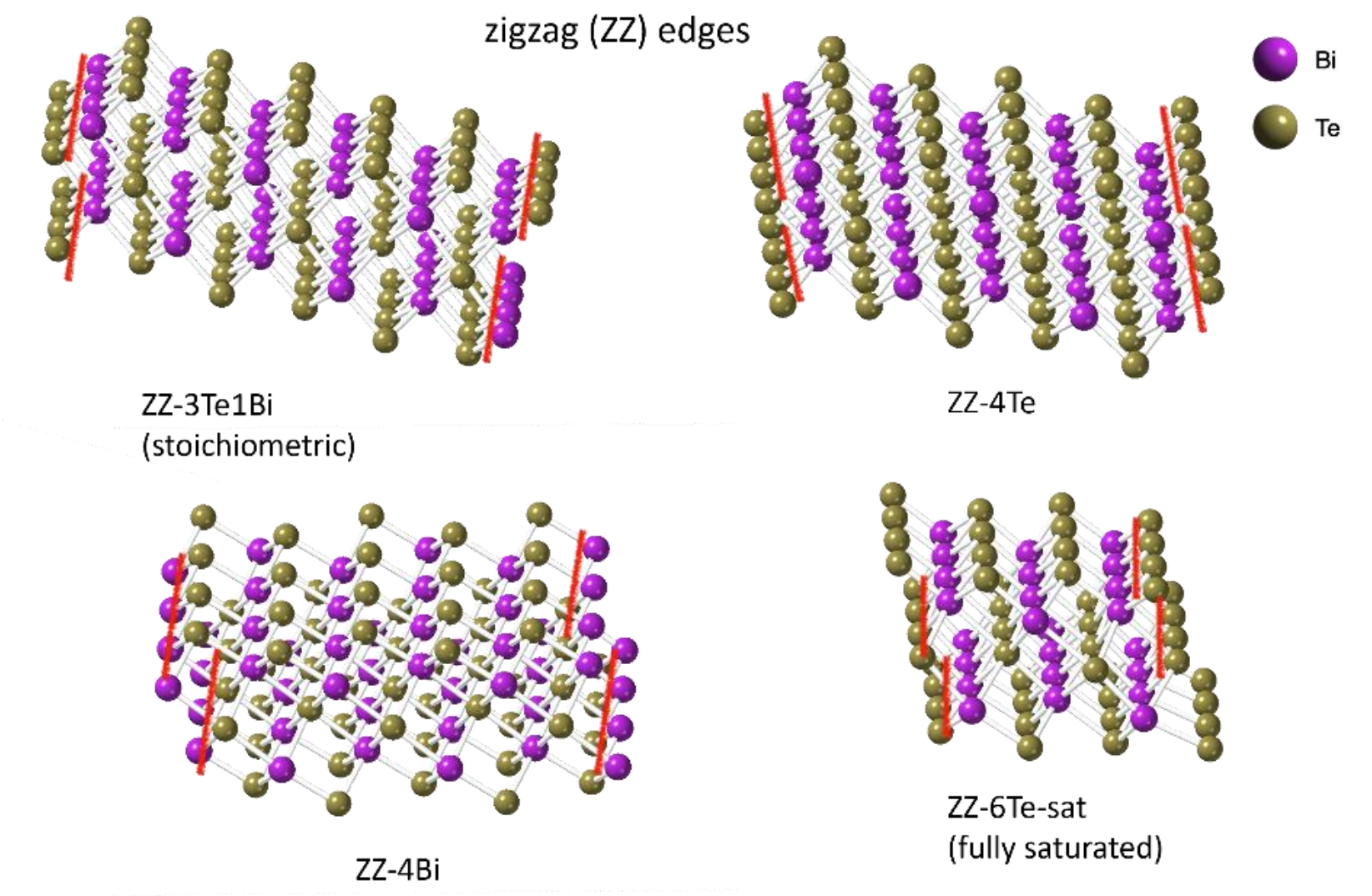

**Figure S10.** Configuration of various zigzag (ZZ) edges in $Bi_2Te_3$ investigated in the DFT framework

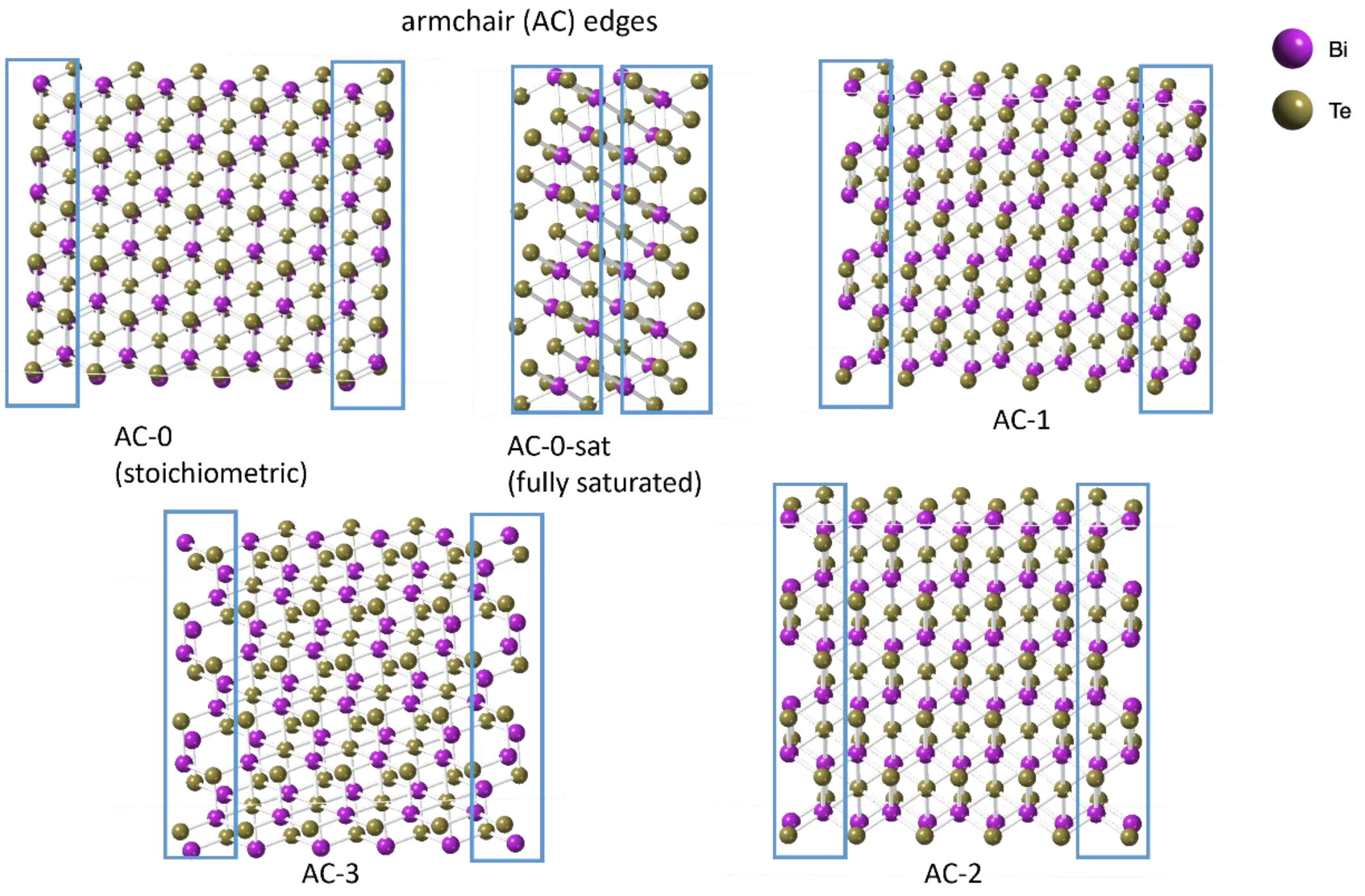

**Figure S11.** Configuration of various armchair edges in Bi2Te3 investigated in the DFT framework

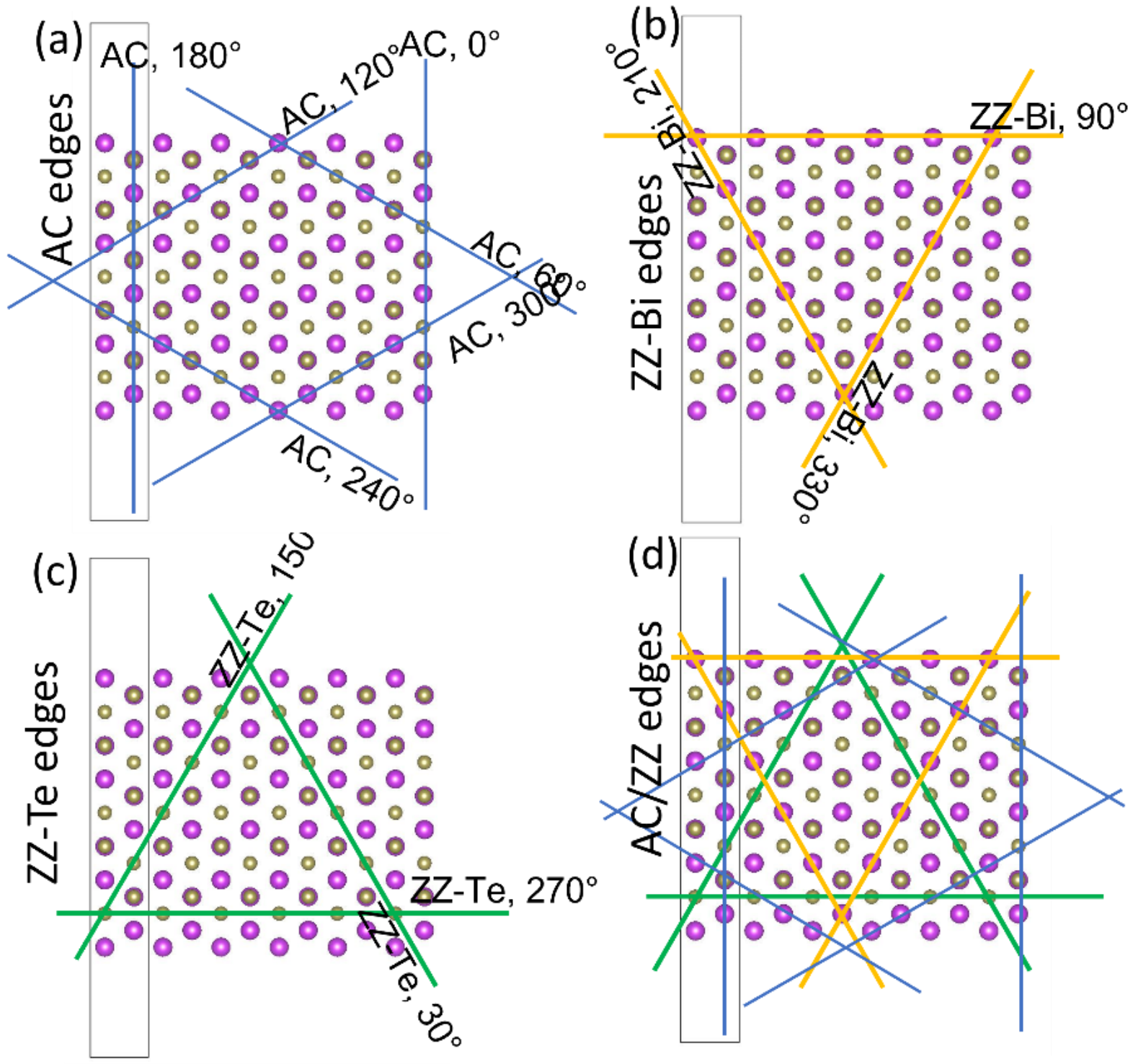

(e)

| Edges | Orientation (°) |
|---|---|
| AC-0 | 0, 60, 120, 180, 240, 300 |
| AC-0-sat | |
| AC-1 | |
| AC-2 | |
| AC-3 | |
| ZZ-4Te | 30, 150, 270 |
| ZZ-3Te1Bi | |
| ZZ-6Te-sat | |
| ZZ-4Bi | 90, 210, 330 |

**Figure S12. (**a) Orientation of AC edges, (b) Orientation of ZZ-Bi edges, (c) Orientation of ZZ-Te edges, (d) Orientation of AC/ZZ edges, (e) Summary table of edge type vs. orientations.